\pdfoutput=1
\documentclass[journal]{IEEEtran}
\ifCLASSOPTIONcompsoc
\usepackage[nocompress]{cite}
\usepackage{mathtools}
\else
\usepackage{cite}
\fi

\ifCLASSINFOpdf
\usepackage{graphicx}
\else
\fi
\newcommand{\aiFigshare}{}
\newcommand{\aiIEEE}{}
\newcommand{\faGithubAlt}{}
\usepackage{float}
\usepackage{makecell}
\usepackage{booktabs}
\usepackage{tabularx}
\usepackage{multirow}
\usepackage{makecell}
\usepackage{amsmath,amssymb,amsfonts}
\usepackage{bm}
\usepackage[noend]{algpseudocode}
\usepackage{algorithmicx,algorithm}
\usepackage{arydshln} 
\usepackage{booktabs}
\usepackage{multirow}
\usepackage{color}
\usepackage{siunitx}
\usepackage{threeparttable}
\usepackage{caption}
\usepackage{placeins}
\usepackage{tikz,xcolor,hyperref}% Make Orcid icon
\usepackage{pgfplots}
\usepackage{verbatim} % 在导言区引入宏包
\colorlet{red}{black}
\definecolor{blue}{rgb}{0,0,0}
\pgfplotsset{compat=1.18}
\definecolor{lime}{HTML}{A6CE39}
\DeclareRobustCommand{\orcidicon}{%
    \begin{tikzpicture}
    \draw[lime, fill=lime] (0,0) 
    circle [radius=0.16] 
    node[white] {{\fontfamily{qag}\selectfont \tiny ID}};    \draw[white, fill=white] (-0.0625,0.095) 
    circle [radius=0.007];    \end{tikzpicture}
    \hspace{-2mm}}
\foreach \x in {A, ..., Z}{%
    \expandafter\xdef\csname orcid\x\endcsname{\noexpand\href{https://orcid.org/\csname orcidauthor\x\endcsname}{\noexpand\orcidicon}}
    }
\hypersetup{hidelinks}

\begin{document}
\title{Memo2496: Expert-Annotated Dataset and Dual-view Adaptive Framework for Music Emotion Recognition}
\author{\IEEEauthorblockN{          
            Qilin Li\orcidA{},~\IEEEmembership{Member,~IEEE,}
            C. L. Philip Chen\orcidC{},~\IEEEmembership{Life Fellow,~IEEE,}
            Tong~Zhang\orcidB{},~\IEEEmembership{Senior Member,~IEEE,}
			}
\thanks{This work was funded in part by the National Natural Science Foundation of China grant under number 62222603, in part by the STI2030-Major Projects grant from the Ministry of Science and Technology of the People's Republic of China under number 2021ZD0200700, in part by the Key-Area Research and Development Program of Guangdong Province under number 2023B0303030001, in part by the Program for Guangdong Introducing Innovative and Entrepreneurial Teams (2019ZT08X214), and in part by the Science and Technology Program of Guangzhou under number 2024A04J6310.}
		\thanks{Qilin Li, C. L. Philip Chen and Tong Zhang are with the Guangdong Provincial Key Laboratory of AI Large Model and Intelligent Cognition, the School of Computer Science and Engineering, South China University of Technology, Guangzhou 510006, China, and with Research Centre for AI Large Models and Intelligent
        Cognition, Pazhou Lab, Guangzhou 510335, China, and also with
        Engineering Research Centre of the Ministry of Education on Health
        Intelligent Perception and Paralleled Digital-Human, South China University
        of Technology, Guangzhou 510006, China. Qilin Li is also with the Department of Automation, Tsinghua University, Beijing 100084, China.

        Tong Zhang is the corresponding author (e-mail: tony@scut.edu.cn).
        
        Dataset and code are available at \aiFigshare\  \href{https://figshare.com/articles/dataset/Memo2496/25827034}{figshare},  \aiIEEE \href{https://dx.doi.org/10.21227/3824-wy49}{ IEEE DataPort} and\\ \faGithubAlt\  \href{https://github.com/QilinLi147/DAMER}{github.com/QilinLi147/DAMER}.}}
\markboth{IEEE Transactions on Affective Computing}
{Li \MakeLowercase{\textit{et al.}}: Memo2496 Dataset and Dual-view Adaptive Framework for MER}

\maketitle

\begin{abstract}
\textcolor{red}{Music Emotion Recognition (MER) is constrained by limited expert annotations and by the need to establish model robustness across heterogeneous corpora. To address these limitations, Memo2496 supplies a reproducible data resource annotated by experts, whilst DAMER provides a general MER framework evaluated on Memo2496 and two established external datasets.} Memo2496, a large-scale dataset, offers 2496 instrumental music tracks with continuous valence-arousal labels, annotated by 30 certified music specialists. \textcolor{red}{Annotation quality is supported by interface familiarisation and duplicate track intra annotator calibration in a normalised circular valence arousal domain.} Furthermore, the Dual-view Adaptive Music Emotion Recogniser (DAMER) is introduced. DAMER integrates three synergistic modules: Dual-Stream Attention Fusion (DSAF) facilitates token-level bidirectional interaction between Mel spectrograms and cochleagrams via cross-attention mechanisms; Progressive Confidence Labelling (PCL) generates reliable pseudo-labels employing curriculum-based temperature scheduling and consistency quantification using Jensen-Shannon divergence; \textcolor{red}{and Style-Anchored Memory Learning (SAML) maintains a labelled contrastive memory queue that regularises the dispersion of embeddings with the same emotion label across acoustically varied samples.} \textcolor{blue}{The primary evaluation follows the binary MER protocol used on PMEmo and 1000songs for cross-dataset comparability, whilst a supplementary continuous regression study on Memo2496 demonstrates direct use of the released segment-level valence and arousal scores.} \textcolor{red}{Extensive experiments on Memo2496, 1000songs, and PMEmo show that DAMER achieves the highest arousal accuracy among the compared methods on Memo2496 and 1000songs and the highest valence accuracy on PMEmo, whilst remaining competitive for PMEmo arousal.} Ablation studies and supplementary diagnostic analyses validate each module's contribution. Both the dataset and source code are publicly available.
\end{abstract}


\section{Introduction}
\IEEEPARstart{M}{usic} Emotion Recognition (MER), a pivotal domain within affective computing, addresses the challenge of automatically identifying and quantifying emotional content conveyed or evoked by musical stimuli \textcolor{red}{\cite{9229494,10.1145/2168752.2168754,kang2025survey,eerola2026meta}}. Distinguished from speech or visual emotion recognition, MER encapsulates affective information hierarchically, from micro-level acoustic features to macro-level compositional elements. This presents unique modelling challenges, necessitating sophisticated temporal and multimodal processing strategies \cite{8327886,10735097}. \textcolor{red}{MER supports affect-aware music retrieval and recommendation, and its representations also inform emotion-conditioned music generation and music-intervention systems \cite{10.1145/2168752.2168754,9745312,10345728,10963042}.} Positioned across music information retrieval, signal processing, machine learning, and cognitive science, MER's methodological progression from handcrafted feature extraction to end-to-end deep representation learning and multimodal fusion mirrors broader trends in affective computing, rendering its advances transferable to adjacent domains.

\textcolor{red}{MER remains difficult to generalise because emotion annotations are subjective and acoustic-emotion mappings vary across datasets and cultures \cite{kang2025survey,watcharasupat2025uncertainty,lee2025globalmood}. Individual listener factors further motivate personalised modelling \cite{xu2021effects,zhang2025personalised}. Existing public corpora differ in scale, repertoire, modality, and annotation protocol \cite{10.1145/2506364.2506365,7178058,Zhang:2018:PDM:3206025.3206037,ferreira_ismir_2019,hung2021emopia,gomez2023trompa}. PMEmo contains 794 pop songs, EMOPIA contains 1,087 clips from 387 pop-piano songs, and the original VGMIDI publication reports 95 emotion-annotated piano arrangements of video-game music \cite{Zhang:2018:PDM:3206025.3206037,hung2021emopia,ferreira_ismir_2019}. Generic consistency-based pseudo-labelling and cross-cultural domain adaptation provide useful foundations, but neither directly addresses the joint problem of uncertain labels, complementary acoustic views, and acoustic variability between tracks in MER \cite{sohn2020fixmatch,chen2018crosscultural}.}

\textcolor{red}{Memo2496 and DAMER address connected limitations in reproducible MER. Memo2496 provides a data resource annotated by experts for reliable evaluation. DAMER provides a general acoustic learning framework that addresses complementary representations, uncertain pseudo labels, and variation across tracks. The framework does not depend on metadata or annotation fields specific to Memo2496, and its architecture and learning objective apply independently to Memo2496, 1000songs, and PMEmo. The principal contributions are summarised as follows:}
\begin{enumerate}[\IEEEsetlabelwidth{12}]
    \item \textcolor{red}{Memo2496 provides 2,496 instrumental tracks with continuous valence and arousal annotations from certified music specialists. Its calibration, quality control, and public release procedures support reproducible MER research.}

    \item \textcolor{red}{DAMER provides a learning framework based on two acoustic views that is independent of a particular corpus. DSAF integrates evidence from Mel spectrograms and cochleagrams, PCL controls pseudo label reliability, and SAML regularises representation variation across tracks.}

    \item \textcolor{red}{DAMER is evaluated against established MER approaches and complementary general audio encoders on Memo2496, 1000songs, and PMEmo. The supplementary regression experiment examines direct use of the continuous Memo2496 labels.}
\end{enumerate}

\section{Related Work}
\begin{table*}[!htbp]
    \renewcommand{\arraystretch}{1.1}
    \fontsize{8.2pt}{9.4pt}\selectfont
    \centering
    \captionsetup{justification=centering, font=small}
    \caption{Comparisons of Popular Music Emotion Datasets and Memo2496}
    \label{tab:datasets}
    \begin{tabularx}{\textwidth}{@{}l l l X l@{}}
    \hline
    Dataset & \textcolor{red}{Dataset size} & \textcolor{red}{Labels or signals} & \textcolor{red}{Public audio availability} & \textcolor{red}{Annotation source} \\ \hline
    1000songs\cite{10.1145/2506364.2506365} & 1000 & Valence, Arousal & $\checkmark$ & Crowdsourcing\\
    AMG1608\cite{7178058} & 1608 & Valence, Arousal & $\times$ & Crowdsourcing\\
    VGMIDI\cite{ferreira_ismir_2019} & \textcolor{red}{95 labelled MIDI pieces, original release} & \textcolor{red}{Valence and arousal} & \textcolor{red}{$\times$} & \textcolor{red}{30 crowd annotators}\\
    EMOPIA\cite{hung2021emopia} & \textcolor{red}{1,087 clips from 387 songs} & \textcolor{red}{Russell's four quadrants} & \textcolor{red}{YouTube IDs and timestamps} & \textcolor{red}{Four dedicated annotators}\\
    TROMPA-MER\cite{gomez2023trompa}& 1161 & Valence, Arousal & Downloadable & Public contributors\\
    PMEmo\cite{Zhang:2018:PDM:3206025.3206037} & \textcolor{red}{794 songs with released chorus excerpts} & \textcolor{red}{Valence, arousal, and EDA} & $\checkmark$ & \textcolor{red}{457 participants} \\\hdashline
    \textbf{Memo2496} & \textbf{2496} & \textbf{Valence, Arousal} & $\checkmark$ & \textbf{Certified Music Specialists} \\ \hline
    \end{tabularx}
    \end{table*}

\subsection{Music Emotion Datasets}
\textcolor{red}{MER datasets differ in licensing, emotion taxonomy, annotation granularity, and annotator population, which complicates direct cross-study comparison \cite{kang2025survey,eerola2026meta}. The 1000 Songs Database provides crowdsourced valence-arousal labels for 1,000 music excerpts \cite{10.1145/2506364.2506365}, whilst AMG1608 provides frame-level annotations for 1,608 excerpts \cite{7178058}. These datasets improve scale but require explicit treatment of annotator agreement and corpus-specific label distributions.}

\textcolor{red}{PMEmo provides 794 pop songs with valence-arousal annotations and simultaneous electrodermal activity signals \cite{Zhang:2018:PDM:3206025.3206037}. EMOPIA provides 1,087 clips segmented from 387 pop-piano songs and assigns clip-level labels using Russell's four quadrants \cite{hung2021emopia}. Recent resources expand this landscape: GlobalMood evaluates 1,180 songs from 59 countries through large-scale annotation across five cultural settings \cite{lee2025globalmood}, whilst MERGE provides quality-controlled audio and audio-lyrics variants for static MER \cite{louro2026merge}. Their repertoire, modality, and annotation protocols differ substantially. Table~\ref{tab:datasets} therefore reports dataset units and annotation sources explicitly.}

\subsection{Music Emotion Recognition Methods}
\textcolor{red}{Traditional MER combines handcrafted timbral, rhythmic, spectral, and chroma descriptors with SVM or SVR classifiers \cite{kim2010music}. Pouyanfar and Sameti organise SVM classifiers in a two-level hierarchy for four-class MER \cite{pouyanfar2014twolevel}. Deep MER models combine convolutional and recurrent processing of spectrograms \cite{du2020cnnbilstm}, whilst Transformer encoders provide global audio representations \cite{gong2021ast}. Recent work uses dual-scale attention for personalised dynamic prediction \cite{zhang2025personalised}, and percussion and instrumentation features show that non-melodic acoustic cues remain informative \cite{redinho2026percussion}. Data efficiency, annotation subjectivity, and cross-corpus generalisation remain important limitations.}

\textcolor{red}{Cross-cultural MER uses adversarial domain adaptation to transfer a model trained on Western pop music to Chinese pop music \cite{chen2018crosscultural}. Complementary MER approaches fuse audio and lyrics \cite{quilingking2020multimodal}, symbolic and acoustic features \cite{10317539}, or heterogeneous descriptors through stacked ensembles \cite{chen2020multimodal}. These approaches address specific sources of variation, but reliable labels and stable styles are often assumed.}

\textcolor{blue}{Consistency-based pseudo-labelling methods such as FixMatch \cite{sohn2020fixmatch} and contrastive learning frameworks such as supervised contrastive learning \cite{khosla2020supervised} and Momentum Contrast \cite{he2020momentum} provide important general principles.} \textcolor{red}{Recent MER studies also demonstrate both the difficulty of estimating dispersion in subjective valence-arousal responses \cite{watcharasupat2025uncertainty} and the value of combining semi-supervised graph learning with contrastive regularisation \cite{peintner2025srgnn}. However, these methods are not designed for the joint presence of cross-view acoustic complementarity, subjective emotion ambiguity, and acoustic variability between tracks in music emotion recognition. This motivates a task-specific integration of complementary representations, reliability-aware pseudo-labelling, and labelled contrastive memory.}
\section{Memo2496 DATASET}
\subsection{Data Collection}
\textcolor{red}{Memo2496 comprises 2,496 CC0 instrumental tracks collected primarily from Pixabay (1,507; 60.38\%) and FreePD (988; 39.58\%), with one track from Storyblocks (0.04\%). Artist, album, genre, and related tags for individual tracks are removed before annotation to prevent semantic information leakage and are not retained as complete controlled variables. Complete verified genre labels are unavailable; the source counts describe provenance rather than genre, and no claim of genre balance or genre coverage is made.} \textcolor{red}{Instrumental-only selection targets acoustic emotion: lyrics introduce an additional semantic channel requiring multimodal modelling \cite{malheiro2016bi,9084846}, whilst perceived responses may vary across listeners \cite{xu2021effects}.} Tracks are transcoded to $44.1$ kHz, $16$-bit stereo, $1411$ kbps. \textcolor{blue}{Spleeter-based vocal screening rejects tracks with non-negligible singing; survivors are manually verified.} Loudness is normalised to $-23$ LUFS (EBU R128), and duration filtering retains $30$-$300$ s excerpts.

Thirty specialists from \textbf{South China University of Technology} ($12$ male, $18$ female; aged $18$-$28$, mean $24$) performed annotation. All hold formal music-theory training and instrumental proficiency, with backgrounds spanning piano, traditional Chinese instruments, violin, and others. \textcolor{red}{Recent MER evidence shows that annotation design and response uncertainty affect label reliability \cite{kang2025survey,watcharasupat2025uncertainty}. The specialist cohort is therefore paired with calibration and cross annotation to support annotation quality; expertise alone is not treated as sufficient evidence of reliability.}
\subsection{Annotation Protocol}
\begin{figure*}[t]
\centering
\includegraphics[width=\linewidth]{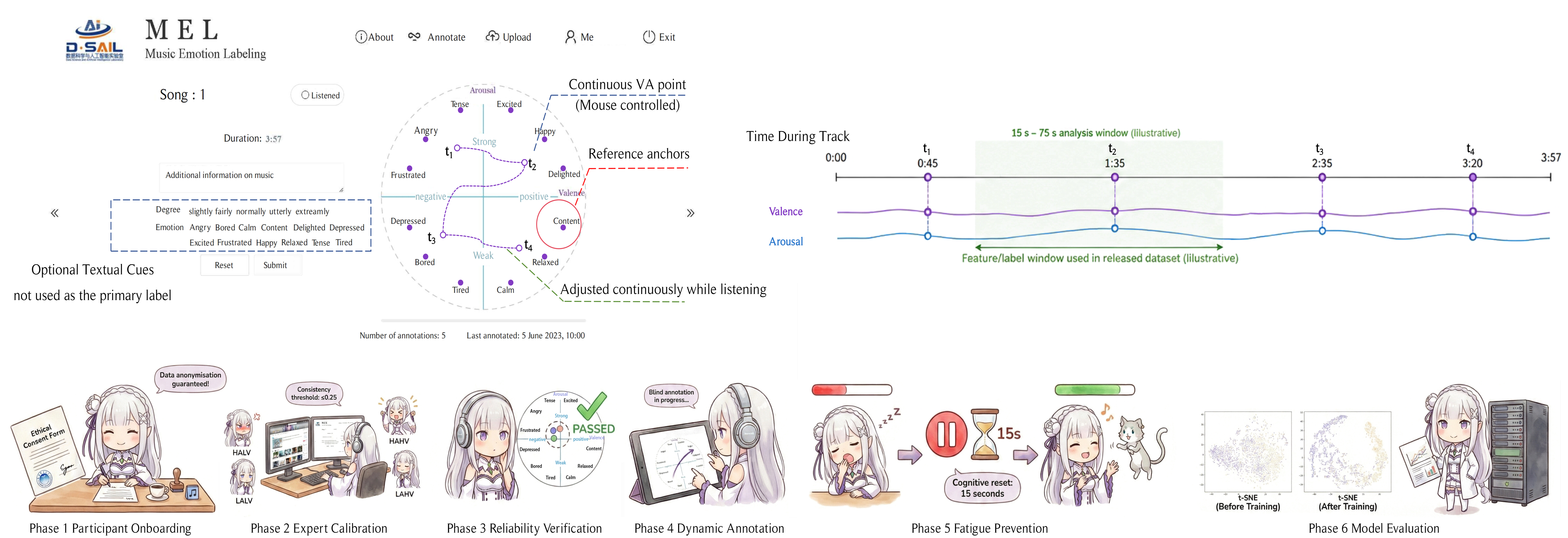}
\caption{\textcolor{blue}{Memo2496 annotation protocol and Mel labelling interface.
\textbf{(Top)} Web-based circumplex interface: annotators continuously adjust a valence-arousal point with a mouse, touchpad, or touchscreen during playback; the score-annotation module records a time-series trajectory; twelve ring labels are spatial reference anchors only; the left panel lists optional supplementary cues; the timeline shows recorded valence and arousal series (15-75~s window illustrative).
\textbf{(Bottom)} End-to-end annotation experiment protocol (consent, training, calibration, formal labelling, and rest intervals).}}
\label{fig:annotation_protocol_composite}
\end{figure*}

\textcolor{blue}{\textcolor{red}{Musical emotions are represented as continuous valence and arousal coordinates on a circular domain grounded in Russell's circumplex model of affect \cite{russell1980circumplex}.} During listening, annotators use a mouse, touchpad, or touchscreen to position and move a point on the circumplex disc shown in Fig.~\ref{fig:annotation_protocol_composite} (top panel), adjusting coordinates in real time as the perceived emotion evolves. The score-annotation module records these positions as a time-series trajectory rather than a single static click. Twelve emotion terms (e.g., Angry, Calm, Happy) appear around the disc at theoretically motivated locations; they serve as spatial reference anchors that partition the plane for interpretation, not as mutually exclusive response categories. The full experiment protocol is summarised in the lower panel of the same figure.}

\textcolor{blue}{The left panel of the interface lists optional degree modifiers and emotion words. After listening, annotators may enter brief textual notes guided by these cues; such entries are supplementary and do not replace the continuous valence-arousal trajectory stored by the system. Perceived intensity is quantified by Euclidean distance from the origin on the circumplex plane; real-time tracking with spatial reference anchors supports labelling precision. Before formal annotation, annotators complete a familiarisation and calibration stage. The familiarisation stage introduces the annotation interface and the valence-arousal space through preset clips sampled from four extreme affective regions. This stage is designed to ensure operational understanding of the task rather than to impose a preferred label interpretation. After familiarisation, annotators evaluate 20 practice tracks, including one covert duplicate, for entry calibration. \textcolor{red}{The covert duplicate assesses short term intra annotator consistency when operating the continuous annotation tool. Calibration is successful when the mean Euclidean distance between its two continuous annotations does not exceed $0.25$. Annotators who do not satisfy this criterion receive further instruction before formal annotation. The value $0.25$ is a study specific tolerance for Memo2496 rather than a universal standard. The normalised circular valence arousal domain has radius $1$ and diameter $2$, so the tolerance equals $25\%$ of the radius and $12.5\%$ of the diameter. When valence and arousal discrepancies are equal, each component is $0.25/\sqrt{2}=0.177$, or $8.84\%$ of one axis span. AMG1608 and PMEmo establish duplicate based consistency screening as an annotation quality control procedure \cite{7178058,Zhang:2018:PDM:3206025.3206037}. Continuous music emotion test retest research reports within participant differences of approximately $8\%$ and average rating dispersion of $12.2\%$ of the scale range \cite{doi:10.1177/0305735611430079}. These studies support the screening procedure and scale relative magnitude, but they do not determine the exact Memo2496 tolerance.}} Annotation accuracy is maintained by a structured rest protocol, including a $15$-second cognitive reset interval following each track, mitigating carryover effects. It is crucial to note that the annotations are based on the perceived emotions conveyed by the music, rather than the annotators' personal emotional responses \cite{gabrielsson2001emotion,schubert2013emotion}. Daily sessions are capped at one hour. The $2496$ tracks are divided into three groups and 30 annotators into three cohorts. A cross-annotation strategy assigns each track group to two distinct cohorts, ensuring each annotator labels $1664$ tracks. This distributes workload and enables inter-annotator agreement assessment.

\textcolor{blue}{\paragraph{From trajectories to released segment labels.} Raw annotations are stored in CSV files keyed by \texttt{song\_id} and temporal markers, with continuous coordinates in $[-1,1]$ for both dimensions and segment-wise standard deviations when multiple samples exist within a window. Trajectory samples are recorded synchronously with playback progress. For model release, the \texttt{label\_process} utility aligns trajectories with the fixed 15-75~s analysis segment and aggregates coordinates within that window by temporal averaging to obtain segment-level valence and arousal scores; when two expert cohorts annotate the same track group, cohort-wise scores are merged by arithmetic averaging after quality filtering. These continuous segment scores are exported in \texttt{label\_v.npy} and \texttt{label\_a.npy}. Binary evaluation in the main experiments applies a threshold of zero to these scores for cross-dataset comparability; supplementary regression in Section~\ref{subsec:memo2496_regression} uses the continuous values directly.}
\subsection{Feature Extraction}
Mel spectrograms apply pre-emphasis ($x'(n)=x(n)-0.97x(n-1)$), $N$-sample Hamming-windowed frames with $50\%$ overlap ($N/2$ hop), and $2N$-point short-time Fourier transform (STFT); magnitudes pass through a $128$-filter Mel bank, and log-compressed squared band energies form a time $\times$ $128$-band map.

Cochleagrams model peripheral auditory filtering with the same pre-emphasis and framing, an $84$-filter log-spaced Gammatone bank, power-law outer-hair-cell compression, and $\log_{10}$ scaling, yielding $[N,84,87]$ tensors. Both representations are computed on the fixed $15$-$75$~s ($60$~s) excerpt.

\textcolor{blue}{The fixed window from $15$ to $75$ seconds is chosen for two reasons. First, initial orientation studies indicate that listeners often require several seconds before stable emotional judgement emerges \cite{10.1371/journal.pone.0173392,doi:10.1177/0305735611430079}, so discarding the opening segment reduces onset bias. Second, a one-minute interval provides sufficient temporal span for continuous affective development whilst remaining compatible with reliable music emotion annotation protocols \cite{10.1145/2647868.2655019}. These studies therefore support the present design from both perceptual stabilisation and annotation reliability perspectives.}
\textcolor{blue}{Although the retained raw tracks span $30$ to $300$ seconds after pre-processing, model training and inference do not operate on variable-length whole-track audio. Instead, feature extraction is based primarily on a fixed $60$-second window from $15$ to $75$ seconds, so that Mel spectrogram and cochleagram inputs remain temporally comparable across samples. When a track does not fully cover this target window, zero-padding is applied only to the missing portion to preserve a consistent input shape.}

\textcolor{blue}{The released labels used for model training are aligned with these fixed window samples rather than treated as a single static label for an entire raw track. This design is consistent with the continuous valence-arousal annotation protocol and supports learning from controlled affective evidence under a unified input format.}
\begin{figure*}[htbp]
    \centerline{\includegraphics[width=\linewidth]{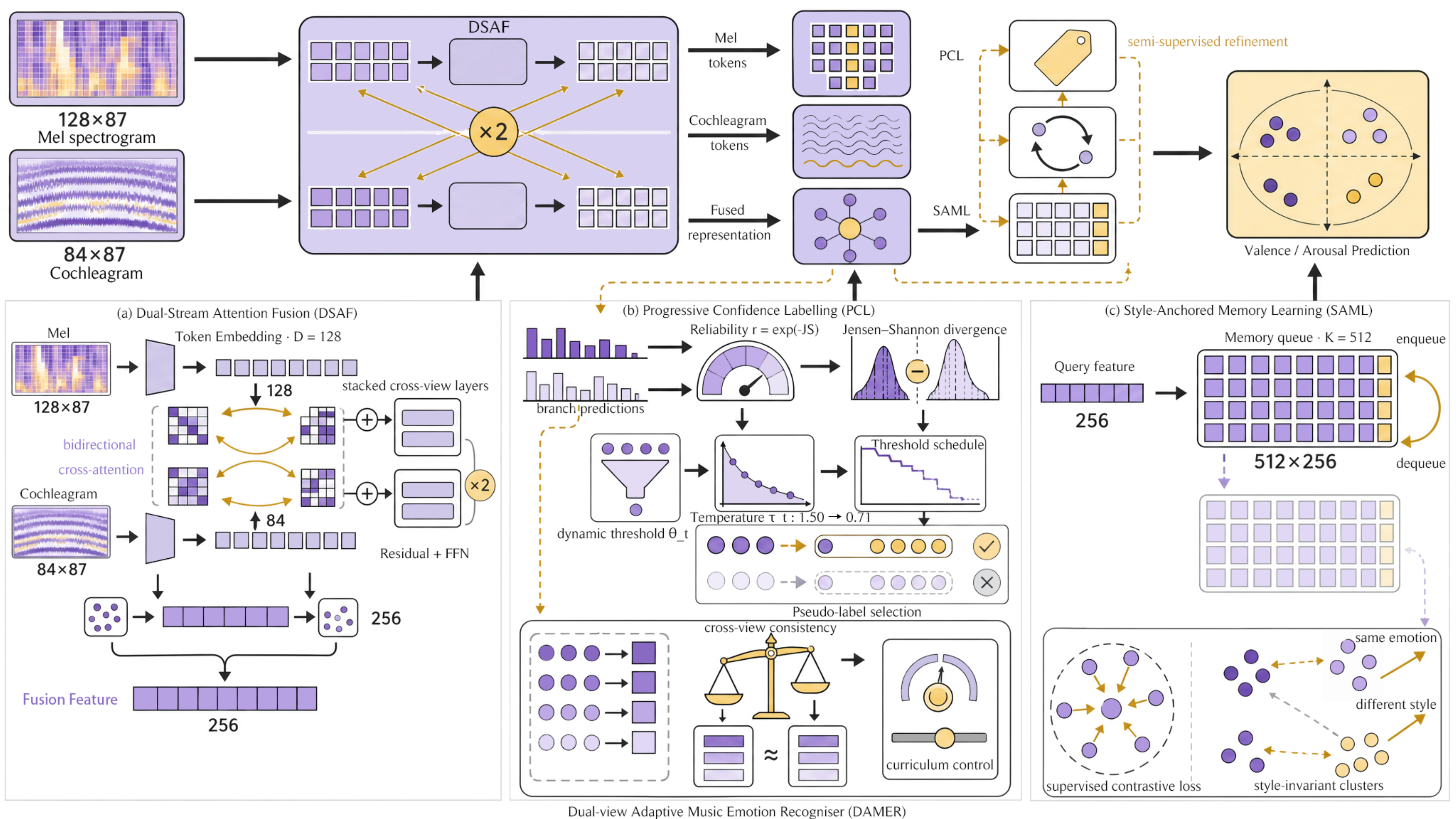}}
    \caption{\textcolor{blue}{The DAMER framework processes complementary Mel spectrogram and cochleagram inputs through DSAF for token-level cross-view fusion, PCL for reliability-aware pseudo-labelling with curriculum temperature control, and SAML for style-anchored contrastive memory learning. Three parallel classification heads support branch-specific and fused predictions for binary valence/arousal recognition.}}
    \label{DAMER}
    \end{figure*}
\section{METHOD IMPLEMENTATION}
\textcolor{red}{DAMER receives Mel spectrogram and cochleagram inputs derived from the same audio segment. The Mel representation preserves the broad spectral envelope and harmonic organisation, whereas the cochleagram provides perceptually motivated frequency resolution and non-linear auditory compression. Emotion cues need not be equally salient in these two views, so the framework retains both representations and allows each to refine the other. Recent MER evidence that percussion and instrumentation contribute beyond predominantly melodic descriptors further supports complementary acoustic modelling \cite{redinho2026percussion}. DAMER does not use a corpus identifier, genre metadata, or any dataset specific annotation field.}

\textcolor{red}{As shown in Fig.~\ref{DAMER}, the method combines three complementary learning functions. DSAF performs token-level interaction between the acoustic views instead of relying on direct feature concatenation. PCL limits the influence of uncertain pseudo labels by combining prediction confidence with cross-view agreement. SAML extends labelled contrastive comparisons beyond the current mini batch and regularises the dispersion of representations with the same emotion label. Mel, cochleagram, and fused classification heads provide branch specific and fused predictions. The same architecture and objectives apply to all three evaluated corpora.}

\subsection{\textcolor{red}{Core Learning Modules}}
\paragraph{\textcolor{red}{Dual-Stream Attention Fusion.}}
\textcolor{red}{The two feature maps differ in their frequency resolution and cannot be compared directly. DSAF first treats their time-frequency elements as token sequences and projects both sequences to the common dimension $D$:}
\begin{align}
{\color{red}\mathbf{h}^{\mathrm{Mel}}_i}
&{\color{red}=\mathbf{W}_{\mathrm{Mel}}\mathbf{x}^{\mathrm{Mel}}_i+\mathbf{b}_{\mathrm{Mel}},} \\
{\color{red}\mathbf{h}^{\mathrm{Coch}}_j}
&{\color{red}=\mathbf{W}_{\mathrm{Coch}}\mathbf{x}^{\mathrm{Coch}}_j+\mathbf{b}_{\mathrm{Coch}}.}
\end{align}
\textcolor{red}{Here, $\mathbf{x}^{\mathrm{Mel}}_i$ and $\mathbf{x}^{\mathrm{Coch}}_j$ denote the input features at the $i$th and $j$th tokens, respectively. The resulting sequences satisfy $\mathbf{H}^{\mathrm{Mel}}\in\mathbb{R}^{B\times N_{\mathrm{Mel}}\times D}$ and $\mathbf{H}^{\mathrm{Coch}}\in\mathbb{R}^{B\times N_{\mathrm{Coch}}\times D}$, where $B$ is the batch size and $N_{\mathrm{Mel}}$ and $N_{\mathrm{Coch}}$ are the respective token counts. This common space supports fine-grained interaction without requiring equal sequence lengths.}

\textcolor{red}{Bidirectional multihead cross attention then exchanges complementary information. Mel tokens query the cochleagram sequence, whilst cochleagram tokens query the Mel sequence:}
\begin{align}
{\color{red}\mathbf{H}_{\mathrm{M}\leftarrow\mathrm{C}}}
&{\color{red}=\operatorname{MHA}(\mathbf{H}^{\mathrm{Mel}},\mathbf{H}^{\mathrm{Coch}},\mathbf{H}^{\mathrm{Coch}}),} \\
{\color{red}\mathbf{H}_{\mathrm{C}\leftarrow\mathrm{M}}}
&{\color{red}=\operatorname{MHA}(\mathbf{H}^{\mathrm{Coch}},\mathbf{H}^{\mathrm{Mel}},\mathbf{H}^{\mathrm{Mel}}).}
\end{align}
\textcolor{red}{In each direction, the querying view preserves its token organisation, whilst the keys and values supply contextual evidence from the complementary view. For each of the $H$ heads, scaled dot product attention is}
\begin{equation}
{\color{red}\operatorname{Attn}(\mathbf{Q},\mathbf{K},\mathbf{V})=
\operatorname{softmax}\!\left(\frac{\mathbf{Q}\mathbf{K}^{\mathsf T}}{\sqrt{d_k}}\right)\mathbf{V}.}
\end{equation}
\textcolor{red}{The term $d_k$ denotes the key dimension of one attention head. The head outputs are concatenated and linearly projected to form the multihead output $\mathbf{A}$. Residual paths retain view-specific evidence, and layer normalisation stabilises the successive exchanges. One Transformer block is summarised by}
\begin{equation}
{\color{red}\begin{aligned}
\mathbf{H}'&=\operatorname{LN}(\mathbf{H}+\operatorname{Drop}(\mathbf{A})),\\
\mathbf{H}''&=\operatorname{LN}(\mathbf{H}'+\operatorname{Drop}(\operatorname{FFN}(\mathbf{H}'))).
\end{aligned}}
\end{equation}
\textcolor{red}{Here, $\operatorname{Drop}$, $\operatorname{LN}$, and $\operatorname{FFN}$ denote dropout, layer normalisation, and a position-wise feed forward network. The feed forward network uses two linear layers with GELU activation and an expansion factor of $4$. After $L$ stacked blocks, average pooling produces one vector for each branch. Their concatenation passes through the fusion projection to obtain the representation used by the fused head and SAML. This arrangement preserves branch-specific predictions for consistency estimation while allowing the fused head to exploit aligned evidence from both views.}

\paragraph{\textcolor{red}{Progressive Confidence Labelling.}}
\textcolor{red}{Pseudo labels derived from uncertain early predictions may reinforce classification errors. PCL controls this confirmation bias through two coupled curricula. The temperature determines prediction sharpness, and the selection threshold determines which unlabelled samples enter the pseudo-labelled subset. Both quantities decrease linearly over $T$ epochs:}
\begin{align}
{\color{red}\tau_t} &{\color{red}=\tau_{\max}-(\tau_{\max}-\tau_{\min})\frac{t}{T},} \label{eq9}\\
{\color{red}\theta_t} &{\color{red}=\theta_0-(\theta_0-\theta_{\min})\frac{t}{T}.} \label{eq16}
\end{align}
\textcolor{red}{The bounds are $\tau_{\max}=1.5$, $\tau_{\min}=0.7$, $\theta_0=0.65$, and $\theta_{\min}=0.35$. The higher initial temperature avoids premature concentration on one class. Its gradual reduction sharpens the distribution as the branch classifiers become more stable. In parallel, the decreasing threshold expands coverage from high-confidence samples to more difficult samples. Temperature-scaled branch probabilities are}
\begin{align}
{\color{red}\mathbf{p}^{\mathrm{Mel}}}
&{\color{red}=\operatorname{softmax}(\mathbf{z}^{\mathrm{Mel}}/\tau_t),} \\
{\color{red}\mathbf{p}^{\mathrm{Coch}}}
&{\color{red}=\operatorname{softmax}(\mathbf{z}^{\mathrm{Coch}}/\tau_t).}
\end{align}
\textcolor{red}{Prediction certainty alone does not reveal whether the two acoustic views support the same decision. PCL therefore averages the two distributions and measures their agreement through Jensen-Shannon divergence. The fused distribution, reliability, and selection confidence are}
\begin{align}
{\color{red}\mathbf{p}^{\mathrm{Fuse}}}
&{\color{red}=\tfrac{1}{2}(\mathbf{p}^{\mathrm{Mel}}+\mathbf{p}^{\mathrm{Coch}}),} \\
{\color{red}r}
&{\color{red}=\exp[-\operatorname{JS}(\mathbf{p}^{\mathrm{Mel}},\mathbf{p}^{\mathrm{Coch}})],} \\
{\color{red}c}
&{\color{red}=r\max(\mathbf{p}^{\mathrm{Fuse}}).}
\end{align}
\textcolor{red}{The Jensen-Shannon divergence is symmetric and bounded within $[0,\log 2]$. Consequently, $r\in[0.5,1]$, with values near unity indicating close agreement between the branch predictions. Multiplication by the maximum fused probability requires both inter-view agreement and class certainty for a high selection score.}

\textcolor{red}{Let $\mathcal{D}_{\mathrm{PL}}=\{i:c_i\geq\theta_t\}$ and $\hat y_i=\arg\max\mathbf{p}^{\mathrm{Fuse}}_i$. The selected pseudo label supervises the cochleagram branch through the reliability-weighted loss}
\begin{equation}
{\color{red}\mathcal{L}_{\mathrm{PL}}=\frac{1}{|\mathcal{D}_{\mathrm{PL}}|}
\sum_{i\in\mathcal{D}_{\mathrm{PL}}}r_i\operatorname{CE}(\mathbf{z}^{\mathrm{Coch}}_i,\hat y_i).}
\end{equation}
\textcolor{red}{Reliability weighting reduces the contribution of a selected sample when the two views remain inconsistent. The coupled schedule therefore changes coverage gradually rather than treating all model predictions as equally trustworthy. The same branch distributions also define the consistency term in the complete objective.}

\paragraph{\textcolor{red}{Style-Anchored Memory Learning and optimisation.}}
\textcolor{red}{Tracks with the same emotion label may differ substantially in instrumentation, production texture, rhythm, and harmonic organisation \cite{athanasopoulos2021harmonic,redinho2026percussion}. Such acoustic variability can disperse their embeddings and weaken the class structure available to a classifier. Representation variability here refers only to the dispersion of embeddings with the same emotion label. It does not denote temporal drift, covariate shift, or a formally labelled style domain. Style-Anchored Memory Learning is the module name and does not indicate the use of style labels or explicit style disentanglement. SAML uses no genre, style, domain, or track identity supervision.}

\textcolor{red}{A first in, first out queue $\mathcal{Q}=\{(\mathbf{k}_j,y_j)\}_{j=1}^{K}$ stores fusion features normalised to unit $\ell_2$ norm together with their emotion labels. After each mini batch, new key-label pairs replace the oldest entries through a circular pointer. This operation provides recent comparisons from several mini batches without introducing a moving-average encoder. For a normalised query $\mathbf{q}_i$ with label $y_i$, the positive set is $\mathcal{P}_i=\{j:y_j=y_i,\ y_j\neq-1\}$. The supervised contrastive loss executed by the implementation is}
\begin{equation}\label{eq:contrastive_compact}
{\color{red}\mathcal{L}_{\mathrm{cont}}=-\frac{1}{N_q}\sum_{i=1}^{N_q}
\log\frac{\sum_{j\in\mathcal{P}_i}\exp(\mathbf{q}_i^{\mathsf T}\mathbf{k}_j/\tau_{\mathrm{cont}})+\epsilon}
{\sum_{k=1}^{K}\exp(\mathbf{q}_i^{\mathsf T}\mathbf{k}_k/\tau_{\mathrm{cont}})+\epsilon}.}
\end{equation}
\textcolor{red}{The numerator aggregates valid queue entries with the same emotion label, whereas the denominator contains every stored key. The contrastive temperature $\tau_{\mathrm{cont}}$ controls the concentration of these similarities, and $\epsilon$ ensures numerical stability. Queries without a positive queue entry contribute zero. The queue therefore extends labelled comparison beyond the current mini batch and encourages compact representations with the same emotion label \cite{khosla2020supervised}. It regularises representation dispersion but does not establish explicit invariance to a measured style variable.}

\textcolor{red}{The complete objective combines the three prediction heads with the PCL and SAML regularisers:}
\begin{equation}\label{l_total}
{\color{red}\mathcal{L}_{\mathrm{total}}=\lambda_1\mathcal{L}_{\mathrm{cls}}+\lambda_2\mathcal{L}_{\mathrm{PL}}+\lambda_3\mathcal{L}_{\mathrm{consistency}}+\lambda_4\mathcal{L}_{\mathrm{cont}},}
\end{equation}
\textcolor{red}{The supervised term $\mathcal{L}_{\mathrm{cls}}$ sums cross entropy over the Mel, cochleagram, and fused prediction heads. The term $\mathcal{L}_{\mathrm{PL}}$ introduces selected pseudo-label supervision, $\mathcal{L}_{\mathrm{consistency}}$ is the Jensen-Shannon divergence between the two branch probabilities, and $\mathcal{L}_{\mathrm{cont}}$ supplies labelled memory regularisation. Their relative contributions are controlled by $\lambda_1$ to $\lambda_4$. The corresponding weights, queue capacity, attention depth, and remaining implementation settings are reported in Section~V to avoid repeating the experimental protocol in the method description.}

\section{Experiments and Analysis}
\subsection{Experimental Implementation}

Experiments are conducted on three publicly available MER datasets: Memo2496, 1000songs, and PMEmo. For all datasets, a stratified 70\%/30\% train/test split is employed to preserve the corpus-specific class proportions across partitions. Both Arousal and Valence dimensions are evaluated independently as binary classification tasks, with labels binarised at the zero threshold. \textcolor{blue}{This binary protocol aligns with PMEmo and 1000songs for cross-dataset comparability with prior MER work. A supplementary continuous regression evaluation on Memo2496 is reported in Section~\ref{subsec:memo2496_regression}.}

\begin{figure*}[t]
    \centering
    \includegraphics[width=0.82\textwidth]{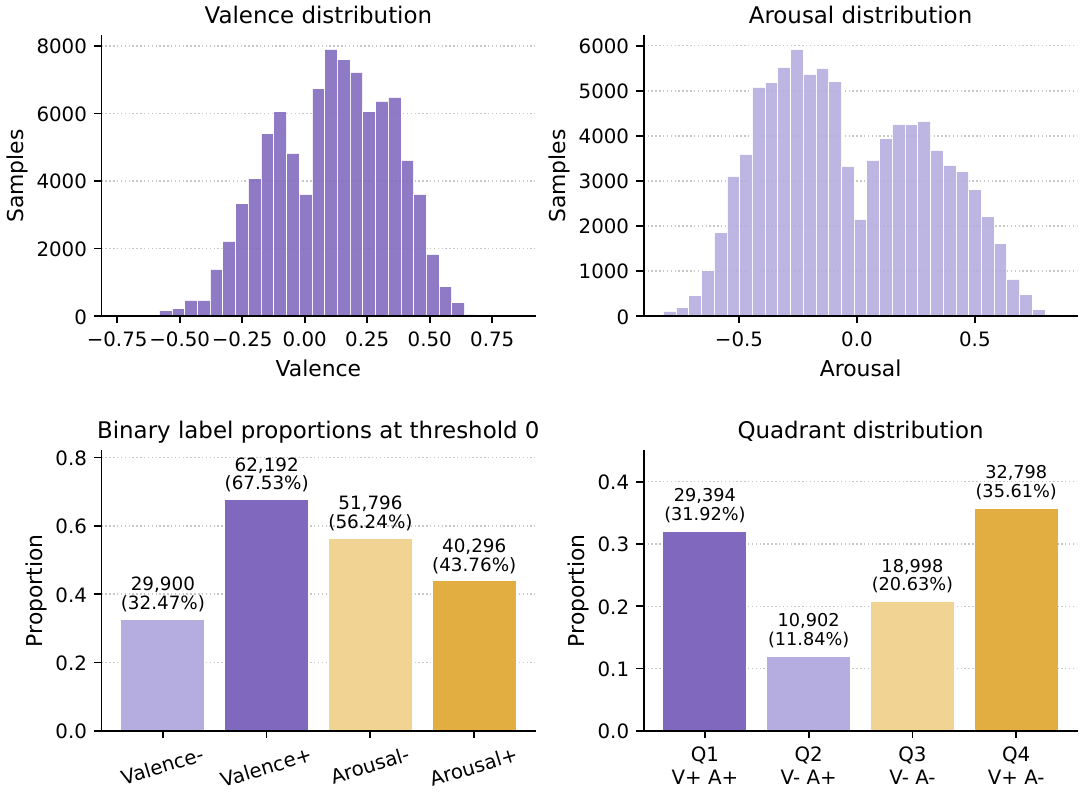}
    \caption{\textcolor{red}{Memo2496 classification label composition. The upper row shows the continuous valence and arousal distributions over the $92{,}092$ classification instances. The lower left panel reports negative and positive counts and proportions after binarisation at $0$, and the lower right panel reports the corresponding quadrant composition. All counts refer to classification instances rather than unique tracks.}}
    \label{fig:label_stats_balance}
\end{figure*}

\textcolor{red}{Fig.~\ref{fig:label_stats_balance} reports the continuous label distributions and the class proportions obtained by binarising the $92{,}092$ classification instances in the released label arrays at $0$. These instances correspond to the inputs used by the classification pipeline and are not counts of unique tracks. Valence has mean $0.1087$ and variance $0.0553$, whereas arousal has mean $-0.0310$ and variance $0.1177$. Valence contains $62{,}192$ positive instances ($67.53\%$) and $29{,}900$ negative instances ($32.47\%$), whilst arousal contains $40{,}296$ positive instances ($43.76\%$) and $51{,}796$ negative instances ($56.24\%$). Neither array contains an exact zero. The valence composition is therefore asymmetric, whereas the arousal composition is closer to even. The means and variances describe central tendency and dispersion only and are not measures of binary class balance. The stratified partition preserves these instance level proportions.}

The DAMER architecture is configured with embedding dimension $D = 128$, fusion dimension $D_f = 256$, $H = 4$ attention heads, and $L = 2$ cross-view transformer layers. \textcolor{red}{The memory queue size is set to $K = 512$.} For instance, $\tau_{\min}=0.7$ and $\tau_{\max}=1.5$, whilst dynamic threshold parameters are $\theta_0 = 0.65$ and $\theta_{\min} = 0.35$. Loss weights are configured as $\lambda_1 = 1.0$, $\lambda_2 = 0.8$, $\lambda_3 = 0.2$, and $\lambda_4 = 0.1$. The model is optimised using Adam with an initial learning rate of $1 \times 10^{-3}$ and weight decay of $1 \times 10^{-4}$, trained for 80 epochs with cosine annealing learning rate scheduling. Mixed-precision training with gradient clipping (maximum norm 5.0) is employed for computational efficiency. The dataset is accessible from \aiFigshare\  Figshare: \href{https://doi.org/10.6084/m9.figshare.25827034}{doi.org/10.6084/m9.figshare.25827034}, \aiIEEE\ IEEE DataPort: \href{https://dx.doi.org/10.21227/3824-wy49}{dx.doi.org/10.21227/3824-wy49} and \faGithubAlt\ \href{https://github.com/QilinLi147/DAMER}{github.com/QilinLi147/DAMER}, with a detailed description, a tutorial on dataset usage, and the DAMER code.\footnote{Direct HTTPS URLs are \url{https://doi.org/10.6084/m9.figshare.25827034}, \url{https://dx.doi.org/10.21227/3824-wy49}, and \url{https://github.com/QilinLi147/DAMER}. The complete dataset can be downloaded through the above channels.} All experiments are conducted on an NVIDIA A100 80GB GPU using PyTorch 1.13.0 with CUDA 12.1. Performance is evaluated using Accuracy (ACC), F1-score, and Area Under the ROC Curve (AUC). \textcolor{red}{For the additional task-specific MER baselines, the same corpus-specific stratified split, temporal crop, dimension-wise zero-threshold binarisation, and evaluation code are used as for DAMER. Model-specific front ends follow the corresponding publications, whilst each output layer is adapted only for binary arousal or valence prediction. Published hyperparameters are retained where compatible, and any necessary selection uses only the training partition.}

\textcolor{blue}{A neighbourhood sensitivity analysis on Memo2496 Arousal shows that the default hyperparameters lie in a stable high-performance region. Across local variations of $\tau_{\min}$, $\tau_{\max}$, $K$, and random seed, the best validation accuracy remains within a narrow interval from $0.8321$ to $0.8372$, which supports the adopted setting as a balanced choice between robustness and computational cost.}

\subsection{\textcolor{blue}{Memo396 Label-Fit Evaluation}}
\begin{figure}[t]
    \centering
    \includegraphics[width=0.92\linewidth]{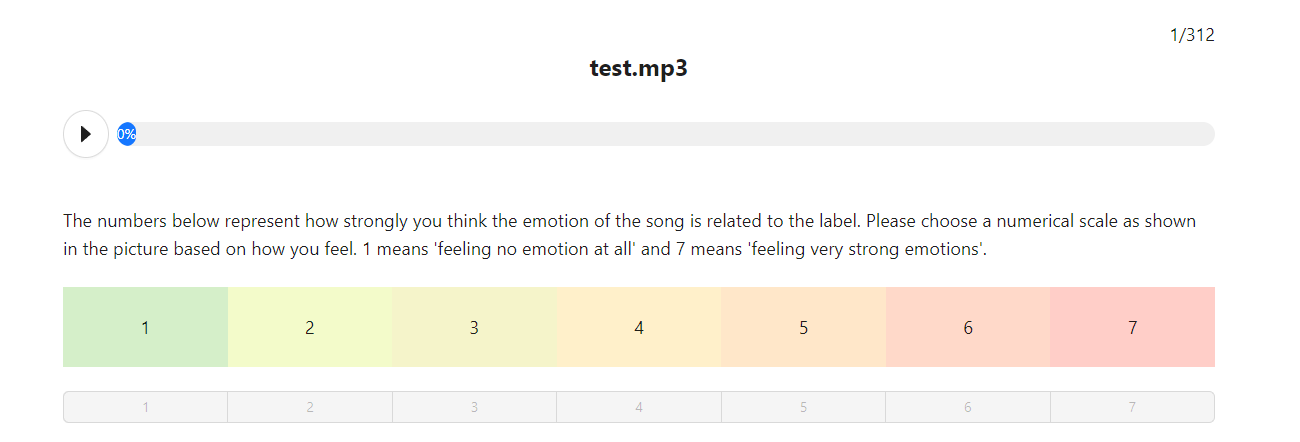}
    \caption{\textcolor{blue}{Interface used in the Memo396 label-fit evaluation. Each clip is scored on a seven-point scale according to how strongly it matches the assigned emotion label.}}
    \label{fig:memo396_interface}
\end{figure}

\textcolor{blue}{The released resources further include the anonymised Memo396 label-fit scoring sheet and category summaries. Memo396 contains $396$ clips from $12$ emotion categories, each scored by $19$ independent raters on a seven-point scale according to how strongly the music matches the assigned label.} \textcolor{red}{The validation panel comprises students and staff members from South China University of Technology, including $10$ women and $9$ men. It consists of $10$ undergraduate students, $5$ postgraduate students, and $4$ postdoctoral researchers. None belongs to the $30$ specialist annotators who annotate Memo2496. The validation raters have no systematic professional music training and are not treated as music experts. Memo396 therefore provides an independent general listener assessment rather than a repeat expert annotation.} \textcolor{blue}{Fig.~\ref{fig:memo396_interface} shows the scoring interface. The overall mean score is $5.1631$, the mean within-track standard deviation is $1.2551$, and Cronbach's alpha is $0.8830$. The highest agreement appears for Happy, Delighted, and Excited, whereas Angry, Depressed, Bored, and Tired remain more ambiguous. These results support the perceptual validity of the released subset whilst also showing that negative categories are harder to match consistently.}

\begin{table}[t]
\renewcommand{\arraystretch}{1.1}
\centering
\captionsetup{justification=centering, font=small}
\caption{\textcolor{red}{Comparison of ACC, F1, and AUC for music emotion recognition on Memo2496 under the reported protocol.}}
\label{CEMemo}
\fontsize{7.5pt}{9pt}\selectfont
\resizebox{\columnwidth}{!}{%
\begin{tabular}{ccccccc}
\hline
\multirow{2}{*}{Model} & \multicolumn{3}{c}{Arousal}                      & \multicolumn{3}{c}{Valence}                      \\ \cline{2-7} 
                       & Acc            & F1             & AUC            & Acc            & F1             & AUC            \\ \hline
SVM\text{*}\cite{kim2010music}                    & 66.83          & 58.34          & 65.96          & 70.48          & 79.64          & 59.69          \\
GCB-net\cite{zhang2019gcbnet}       & 76.83          & 71.43          & 76.24          & 75.62          & 84.34          & 60.20          \\ 
RGCB-net\text{*}\cite{9698117}       & 77.11          & 69.42          & 76.37          & 76.72          & 84.96          & 61.50          \\ 
\textcolor{blue}{CNN-BiLSTM\text{\dag}\cite{du2020cnnbilstm}} & \textcolor{blue}{78.07}          & \textcolor{blue}{72.54}          & \textcolor{blue}{85.23}          & \textcolor{blue}{77.44}          & \textcolor{blue}{86.22}          & \textcolor{blue}{72.16}\\
\textcolor{blue}{FFA-BiGRU\text{\dag}\cite{su2024ffabigru}} & \textcolor{blue}{78.47}          & \textcolor{blue}{74.02}          & \textcolor{blue}{85.64}          & \textcolor{blue}{77.30}          & \textcolor{blue}{85.44}          & \textcolor{blue}{76.66}\\
DSTFN\text{\dag}\cite{he2022music} & 78.04          & 72.41          & 85.91          & 78.21          & 86.24          & 75.34\\
IIOF\cite{chang2024iiof} & 78.14          & 73.28          & 85.33          & 78.36          & 86.44          & 76.81\\
ADFF\cite{huang2022adff} & 75.55 
& 72.77 & 86.03          & 78.19          & 86.89          & 76.91\\
\textcolor{blue}{AST\text{\dag}\cite{gong2021ast}} & \textcolor{blue}{76.71}          & \textcolor{blue}{71.56}          & \textcolor{blue}{84.39}          & \textcolor{blue}{76.66}          & \textcolor{blue}{85.83}          & \textcolor{blue}{72.80}\\
\textcolor{blue}{HuBERT\text{\dag}\cite{sun2025hubertmer}} & \textcolor{blue}{77.65}          & \textcolor{blue}{72.60}          & \textcolor{blue}{85.39}          & \textcolor{blue}{77.16}          & \textcolor{blue}{85.84}          & \textcolor{blue}{74.86}\\
MCGC-net\cite{11162645}       & 79.52 & 72.44 & 78.09 & \textbf{78.61} & \textbf{87.07} & 68.29\\
\textcolor{red}{TPCNet\text{\dag}\cite{hu2026tpcnet}} & \textcolor{red}{81.36} & \textcolor{red}{79.53} & \textcolor{red}{87.82} & \textcolor{red}{77.93} & \textcolor{red}{82.71} & \textcolor{red}{79.87} \\
\textcolor{red}{DE-CNN\text{\dag}\cite{li2024decnn}} & \textcolor{red}{80.19} & \textcolor{red}{78.33} & \textcolor{red}{88.42} & \textcolor{red}{76.80} & \textcolor{red}{80.94} & \textcolor{red}{77.18} \\
\textcolor{red}{Regional 1D-CNN\text{\dag}\cite{shashidhar2023regionalcnn}} & \textcolor{red}{81.11} & \textcolor{red}{79.23} & \textcolor{red}{88.29} & \textcolor{red}{76.74} & \textcolor{red}{83.18} & \textcolor{red}{81.44} \\
\textcolor{red}{Inception-GRU\text{\dag}\cite{han2023inceptiongru}} & \textcolor{red}{80.81} & \textcolor{red}{76.06} & \textcolor{red}{88.63} & \textcolor{red}{76.97} & \textcolor{red}{82.05} & \textcolor{red}{80.19} \\\hdashline
\textbf{DAMER}       & \textbf{82.95 
} & \textbf{80.70} & \textbf{90.21} & 78.34 
 & 84.72 & \textbf{83.31}\\ \hline
\end{tabular}}
\begin{tablenotes}
\item[1] *: official implementation; \dag: reimplementation from published specifications. All methods use the reported corpus protocol.
\end{tablenotes}
\end{table}

\begin{table}[t]
\renewcommand{\arraystretch}{1.1}
\centering
\captionsetup{justification=centering, font=small}
\caption{\textcolor{red}{Comparison of ACC, F1, and AUC for music emotion recognition on 1000songs under the reported protocol.}}
\label{CE1000songs}
\fontsize{7.5pt}{9pt}\selectfont
\resizebox{\columnwidth}{!}{%
\begin{tabular}{ccccccc}
\hline
\multirow{2}{*}{Model} & \multicolumn{3}{c}{Arousal}                      & \multicolumn{3}{c}{Valence}                      \\ \cline{2-7} 
                       & Acc            & F1             & AUC            & Acc            & F1             & AUC            \\ \hline
SVM\text{*}\cite{kim2010music}                    & 59.85          & 64.89          & 60.82          & 53.39          & 53.41          & 58.77          \\
GCB-net\cite{zhang2019gcbnet}       & 77.09          & 83.27          & 76.89          & 67.64          & 76.28          & 63.89      \\    
RGCB-net\text{*}\cite{9698117}       & 78.66          & 83.82          & 77.56          & 68.71          & 77.68          & 64.15      \\    
\textcolor{blue}{CNN-BiLSTM\text{\dag}\cite{du2020cnnbilstm}} & \textcolor{blue}{78.86}          & \textcolor{blue}{83.69}          & \textcolor{blue}{86.54}          & \textcolor{blue}{69.09}          & \textcolor{blue}{78.32}          & \textcolor{blue}{69.62}\\
\textcolor{blue}{FFA-BiGRU\text{\dag}\cite{su2024ffabigru}} & \textcolor{blue}{78.98}          & \textcolor{blue}{83.55}          & \textcolor{blue}{86.92}          & \textcolor{blue}{68.75}          & \textcolor{blue}{77.91}          & \textcolor{blue}{69.52}\\
DSTFN\text{\dag}\cite{he2022music} & 78.94          & 84.58          & 77.69          & 68.71          & 76.95          & 71.40\\
IIOF\cite{chang2024iiof} & 77.03          & \textbf{85.24 }         & 78.82          & 69.91          & \textbf{79.77}          & 73.70\\
ADFF\cite{huang2022adff} & 77.31          & 85.03          & 78.79          & 68.32          & 75.22          & 73.12\\
\textcolor{blue}{AST\text{\dag}\cite{gong2021ast}} & \textcolor{blue}{78.89}          & \textcolor{blue}{83.52}          & \textcolor{blue}{86.33}          & \textcolor{blue}{67.91}          & \textcolor{blue}{76.84}          & \textcolor{blue}{69.31}\\
\textcolor{blue}{HuBERT\text{\dag}\cite{sun2025hubertmer}} & \textcolor{blue}{78.65}          & \textcolor{blue}{83.74}          & \textcolor{blue}{85.61}          & \textcolor{blue}{69.83}          & \textcolor{blue}{78.98}          & \textcolor{blue}{70.46}\\
MCGC-net\cite{11162645}   & 79.03 & 84.49 & 78.21 & 70.05 & 79.42 & 66.59 \\
\textcolor{red}{TPCNet\text{\dag}\cite{hu2026tpcnet}} & \textcolor{red}{80.19} & \textcolor{red}{81.70} & \textcolor{red}{83.91} & \textcolor{red}{67.95} & \textcolor{red}{69.43} & \textcolor{red}{73.84} \\
\textcolor{red}{DE-CNN\text{\dag}\cite{li2024decnn}} & \textcolor{red}{80.89} & \textcolor{red}{81.37} & \textcolor{red}{80.63} & \textcolor{red}{68.56} & \textcolor{red}{68.97} & \textcolor{red}{75.32} \\
\textcolor{red}{Regional 1D-CNN\text{\dag}\cite{shashidhar2023regionalcnn}} & \textcolor{red}{80.94} & \textcolor{red}{81.42} & \textcolor{red}{87.80} & \textcolor{red}{67.92} & \textcolor{red}{69.18} & \textcolor{red}{74.59} \\
\textcolor{red}{Inception-GRU\text{\dag}\cite{han2023inceptiongru}} & \textcolor{red}{79.36} & \textcolor{red}{81.85} & \textcolor{red}{86.93} & \textcolor{red}{68.88} & \textcolor{red}{70.06} & \textcolor{red}{75.05} \\\hdashline
\textbf{DAMER}   & \textbf{81.28} & 82.32 & \textbf{89.15} & \textbf{70.93} & 72.79 & \textbf{77.28} \\ \hline
\end{tabular}}
\end{table}

\begin{table}[t]
\renewcommand{\arraystretch}{1.1}
\centering
\captionsetup{justification=centering, font=small}
\caption{\textcolor{red}{Comparison of ACC, F1, and AUC for music emotion recognition on PMEmo under the reported protocol.}}
\label{CEPMEmo}
\fontsize{7.5pt}{9pt}\selectfont
\resizebox{\columnwidth}{!}{%
\begin{tabular}{ccccccc}
\hline
\multirow{2}{*}{Model} & \multicolumn{3}{c}{Arousal}                      & \multicolumn{3}{c}{Valence}                      \\ \cline{2-7} 
                       & Acc            & F1             & AUC            & Acc            & F1             & AUC            \\ \hline
SVM\text{*}\cite{kim2010music}                     & 49.06          & 59.42          & 53.11          & 53.93          & 64.83          & 50.91          \\
GCB-net\cite{zhang2019gcbnet}        & 83.98   & 90.72   & 71.58   & 73.70   & 84.58   & 62.73          \\ 
RGCB-net\text{*}\cite{9698117}        & 84.76   & 90.60   & 72.33   & 74.32   & 84.86   & 64.21          \\ 
\textcolor{blue}{CNN-BiLSTM\text{\dag}\cite{du2020cnnbilstm}} & \textcolor{blue}{85.23}          & \textcolor{blue}{91.38}          & \textcolor{blue}{84.39}          & \textcolor{blue}{74.83}          & \textcolor{blue}{84.99}          & \textcolor{blue}{74.00}\\
\textcolor{blue}{FFA-BiGRU\text{\dag}\cite{su2024ffabigru}} & \textcolor{blue}{83.40}          & \textcolor{blue}{90.08}          & \textcolor{blue}{84.29}          & \textcolor{blue}{75.35}          & \textcolor{blue}{84.91}          & \textcolor{blue}{75.65}\\
DSTFN\text{\dag}\cite{he2022music} & 83.74          & 90.33          & 83.63          & 75.91          & 85.34          & 76.84\\
IIOF\cite{chang2024iiof} & 84.49          & \textbf{91.96}          & 84.44          & 75.66          & 85.41          & 76.83\\
ADFF\cite{huang2022adff} & 83.76 
& 90.81 & \textcolor{red}{\textbf{84.97}}          & 73.74          & 84.88          & 78.02\\
\textcolor{blue}{AST\text{\dag}\cite{gong2021ast}} & \textcolor{blue}{82.62}          & \textcolor{blue}{89.63}          & \textcolor{blue}{83.01}          & \textcolor{blue}{74.56}          & \textcolor{blue}{84.47}          & \textcolor{blue}{73.76}\\
\textcolor{blue}{HuBERT\text{\dag}\cite{sun2025hubertmer}} & \textcolor{blue}{83.38}          & \textcolor{blue}{90.72}          & \textcolor{blue}{82.82}          & \textcolor{blue}{75.56}          & \textcolor{blue}{84.87}          & \textcolor{blue}{76.41}\\
MCGC-net\cite{11162645}  & 85.81   & 91.42   & 75.12   & 75.42   & 85.11   & 69.81 \\
\textcolor{red}{TPCNet\text{\dag}\cite{hu2026tpcnet}} & \textcolor{red}{\textbf{89.71}} & \textcolor{red}{89.41} & \textcolor{red}{72.86} & \textcolor{red}{73.31} & \textcolor{red}{82.92} & \textcolor{red}{72.82} \\
\textcolor{red}{DE-CNN\text{\dag}\cite{li2024decnn}} & \textcolor{red}{80.49} & \textcolor{red}{90.25} & \textcolor{red}{70.87} & \textcolor{red}{75.53} & \textcolor{red}{80.04} & \textcolor{red}{74.51} \\
\textcolor{red}{Regional 1D-CNN\text{\dag}\cite{shashidhar2023regionalcnn}} & \textcolor{red}{83.03} & \textcolor{red}{88.55} & \textcolor{red}{72.72} & \textcolor{red}{76.44} & \textcolor{red}{83.02} & \textcolor{red}{77.37} \\
\textcolor{red}{Inception-GRU\text{\dag}\cite{han2023inceptiongru}} & \textcolor{red}{80.31} & \textcolor{red}{87.75} & \textcolor{red}{74.94} & \textcolor{red}{72.50} & \textcolor{red}{82.02} & \textcolor{red}{76.39} \\
 \hdashline

\textbf{DAMER}  & \textcolor{red}{85.98}
   & 90.58   & 74.33   & \textbf{77.61}   & \textbf{85.54}   & \textbf{79.71} \\ \hline
\end{tabular}}

\end{table}

\subsection{\textcolor{blue}{Continuous Valence-Arousal Regression on Memo2496}}
\label{subsec:memo2496_regression}

\textcolor{blue}{Although the main comparisons follow the binary MER protocol used on PMEmo and 1000songs for cross-dataset comparability, Memo2496 also provides segment-level continuous valence and arousal scores in \texttt{label\_v.npy} and \texttt{label\_a.npy}, aligned with the fixed 15-75~s feature window. To examine whether the released continuous annotations are directly usable, an additional regression evaluation is conducted on the same Memo2496 split (70/30, seed 42) using the pre-computed Mel and cochleagram inputs. All methods are trained with Huber loss to predict the continuous score, whilst DAMER retains dual-view fusion and disables pseudo-labelling and style-anchored contrastive terms in this setting.} \textcolor{red}{The coefficient of determination is computed independently for each target as $R^2=1-\sum_i(y_i-\hat{y}_i)^2/\sum_i(y_i-\bar{y})^2$ on the same validation partition. Higher $R^2$ indicates that a larger proportion of target variance is explained, whilst negative values indicate performance below the validation-mean predictor. Table~\ref{tab:memo2496_regression} reports RMSE, MAE, Kendall's $\tau$, and $R^2$ under the matched validation protocol. DAMER achieves the highest $R^2$ for arousal ($0.6175$) and valence ($0.4806$), together with the lowest arousal MAE ($0.1556$) and the lowest valence RMSE ($0.1694$) and MAE ($0.1366$). FFA-BiGRU gives the lowest arousal RMSE ($0.2107$), whilst DE-CNN and TPCNet give the highest Kendall's $\tau$ for arousal ($0.5448$) and valence ($0.4421$), respectively. The additional metric therefore supports effective use of the continuous annotations by DAMER, but the results do not indicate uniform superiority on every regression criterion.} \textcolor{blue}{The primary binary evaluation is therefore retained for fair comparison with prior work, whilst the regression study confirms that the dataset contribution is not limited to dichotomised labels. Trajectory-level dynamic prediction is not attempted here and remains a direction for future work using the released continuous annotations.}

\begin{table*}[!t]
\renewcommand{\arraystretch}{1.08}
\centering
\captionsetup{justification=centering, font=small}
\caption{\textcolor{red}{Continuous valence and arousal regression on the Memo2496 validation split under the matched protocol.}}
\label{tab:memo2496_regression}
\fontsize{7.2pt}{8.4pt}\selectfont
\begin{tabular*}{\textwidth}{@{\extracolsep{\fill}}lcccccccc@{}}
\hline
\multirow{2}{*}{Model} & \multicolumn{4}{c}{Arousal} & \multicolumn{4}{c}{Valence} \\ \cline{2-9}
 & RMSE$\downarrow$ & MAE$\downarrow$ & $\tau\uparrow$ & \textcolor{red}{$R^2\uparrow$}
 & RMSE$\downarrow$ & MAE$\downarrow$ & $\tau\uparrow$ & \textcolor{red}{$R^2\uparrow$} \\ \hline
AST\text{\dag}\cite{gong2021ast}
 & 0.2453 & 0.1872 & 0.4987 & \textcolor{red}{0.4893}
 & 0.1897 & 0.1502 & 0.3234 & \textcolor{red}{0.3483} \\
FFA-BiGRU\text{\dag}\cite{su2024ffabigru}
 & \textbf{0.2107} & 0.1621 & 0.5312 & \textcolor{red}{0.5932}
 & 0.1784 & 0.1402 & 0.3956 & \textcolor{red}{0.4237} \\
HuBERT\text{\dag}\cite{sun2025hubertmer}
 & 0.2261 & 0.1732 & 0.5083 & \textcolor{red}{0.5663}
 & 0.1829 & 0.1453 & 0.3448 & \textcolor{red}{0.3944} \\
CNN-BiLSTM\text{\dag}\cite{du2020cnnbilstm}
 & 0.2228 & 0.1699 & 0.4976 & \textcolor{red}{0.5789}
 & 0.1941 & 0.1564 & 0.3070 & \textcolor{red}{0.3180} \\
\textcolor{red}{TPCNet\text{\dag}\cite{hu2026tpcnet}}
 & \textcolor{red}{0.2173} & \textcolor{red}{0.1718} & \textcolor{red}{0.5215} & \textcolor{red}{0.5833}
 & \textcolor{red}{0.1700} & \textcolor{red}{0.1395} & \textcolor{red}{\textbf{0.4421}} & \textcolor{red}{0.4790} \\
\textcolor{red}{DE-CNN\text{\dag}\cite{li2024decnn}}
 & \textcolor{red}{0.2322} & \textcolor{red}{0.1637} & \textcolor{red}{\textbf{0.5448}} & \textcolor{red}{0.5769}
 & \textcolor{red}{0.1855} & \textcolor{red}{0.1646} & \textcolor{red}{0.3914} & \textcolor{red}{0.4646} \\
\textcolor{red}{Regional 1D-CNN\text{\dag}\cite{shashidhar2023regionalcnn}}
 & \textcolor{red}{0.2253} & \textcolor{red}{0.1696} & \textcolor{red}{0.5310} & \textcolor{red}{0.5908}
 & \textcolor{red}{0.1890} & \textcolor{red}{0.1578} & \textcolor{red}{0.3802} & \textcolor{red}{0.4289} \\
\textcolor{red}{Inception-GRU\text{\dag}\cite{han2023inceptiongru}}
 & \textcolor{red}{0.2439} & \textcolor{red}{0.1786} & \textcolor{red}{0.5349} & \textcolor{red}{0.6043}
 & \textcolor{red}{0.1943} & \textcolor{red}{0.1538} & \textcolor{red}{0.3665} & \textcolor{red}{0.4201} \\
\hdashline
\textbf{DAMER}
 & 0.2123 & \textbf{0.1556} & 0.5447 & \textcolor{red}{\textbf{0.6175}}
 & \textbf{0.1694} & \textbf{0.1366} & 0.4400 & \textcolor{red}{\textbf{0.4806}} \\ \hline
\end{tabular*}
\par\vspace{1pt}{\footnotesize \dag: reimplementation under the matched regression protocol. Lower RMSE and MAE and higher $\tau$ and $R^2$ indicate better performance.}
\end{table*}

\begin{figure}[!t]
\centering
\includegraphics[width=\linewidth]{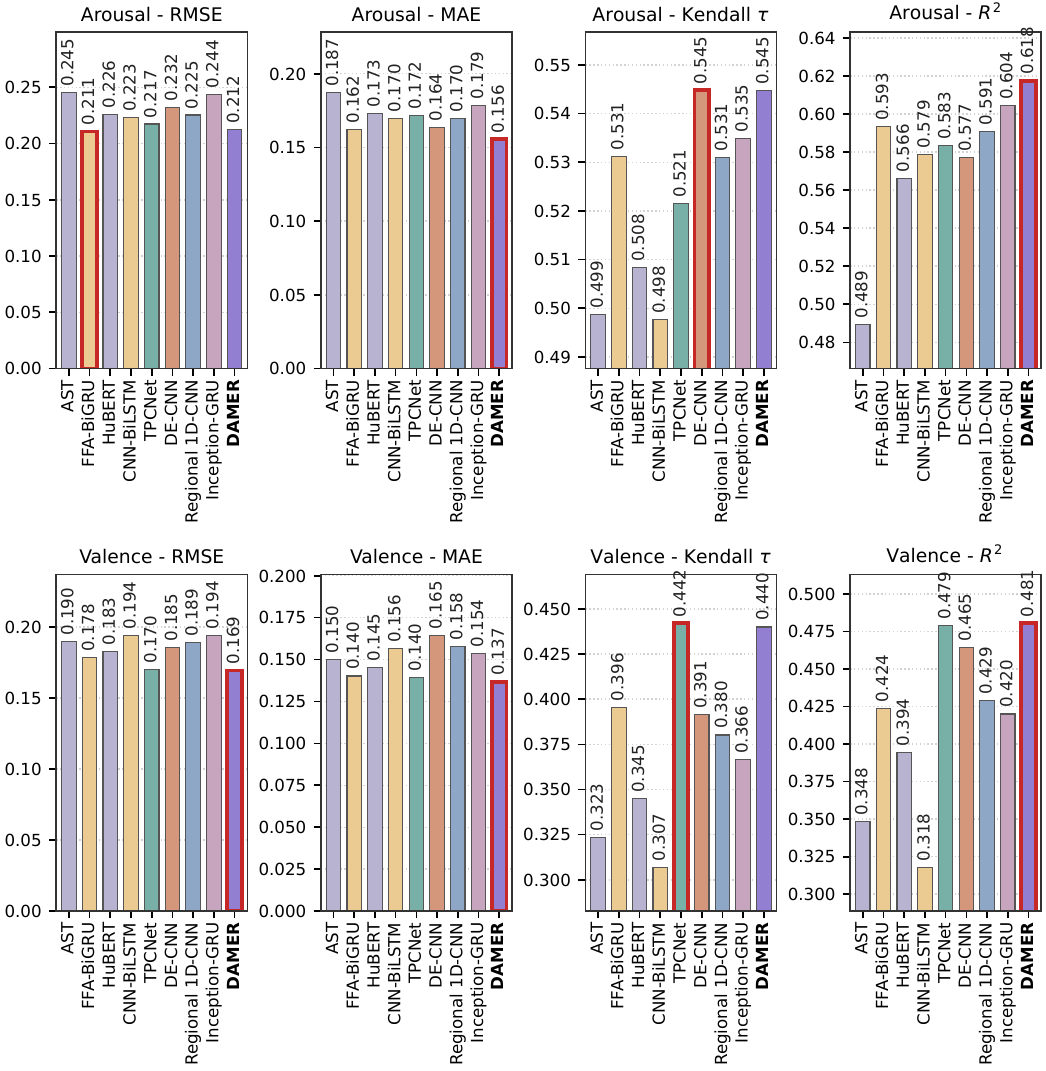}
\caption{\textcolor{red}{Continuous valence and arousal regression on the Memo2496 validation split under the matched protocol. The eight panels report RMSE, MAE, Kendall's $\tau$, and $R^2$ for every method in Table~\ref{tab:memo2496_regression}. Purple bars denote DAMER, and red outlines identify the best result within each panel.}}
\label{fig:memo2496_regression}
\end{figure}

\subsection{Experiments on MER Datasets}

\textcolor{red}{DAMER is compared with fifteen baselines in two groups. The task-specific MER group comprises SVM \cite{kim2010music}, MCGC-net \cite{11162645}, CNN-BiLSTM \cite{du2020cnnbilstm}, FFA-BiGRU \cite{su2024ffabigru}, DSTFN \cite{he2022music}, IIOF \cite{chang2024iiof}, ADFF \cite{huang2022adff}, HuBERT \cite{sun2025hubertmer}, TPCNet \cite{hu2026tpcnet}, DE-CNN \cite{li2024decnn}, Regional 1D-CNN \cite{shashidhar2023regionalcnn}, and Inception-GRU \cite{han2023inceptiongru}. The complementary transfer group contains GCB-net \cite{zhang2019gcbnet}, Residual GCB-net \cite{9698117}, and AST \cite{gong2021ast}. Asterisks identify official implementations, whilst daggers identify reimplementations from published specifications. All methods use the reported corpus protocol.}

\textcolor{red}{Across the expanded comparison, DAMER achieves the highest arousal ACC and AUC on Memo2496 and 1000songs and the highest valence AUC on all three corpora. It also achieves the highest valence ACC and F1 on PMEmo. The strongest PMEmo arousal results are distributed across TPCNet for ACC, IIOF for F1, and ADFF for AUC. These results establish robust cross-corpus performance for DAMER without implying universal superiority for every target and metric.}

\textcolor{red}{The three evaluation corpora differ in repertoire and annotation design. Competitive performance across them supports robustness across corpus conditions rather than a claim of verified genre coverage.}

\begin{figure}[!t]
    \centering
    \includegraphics[width=\linewidth]{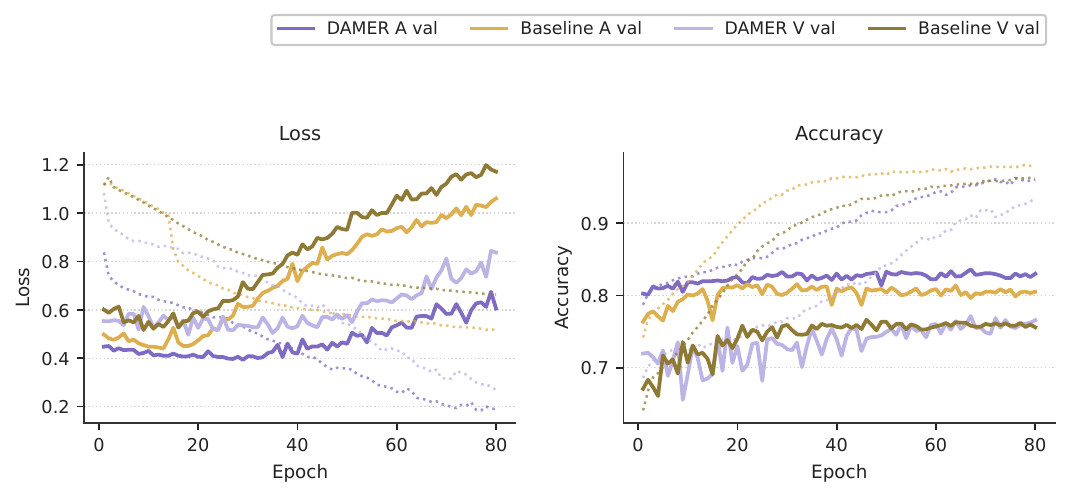}
    \caption{\textcolor{red}{Convergence diagnostics for DAMER and the matched baseline on Memo2496. Solid lines represent validation metrics, whilst dotted lines represent the corresponding training metrics. Colour identifies the model and target dimension.}}
    \label{fig:r21_convergence}
\end{figure}

\begin{figure}[!t]
    \centering
    \includegraphics[width=\linewidth]{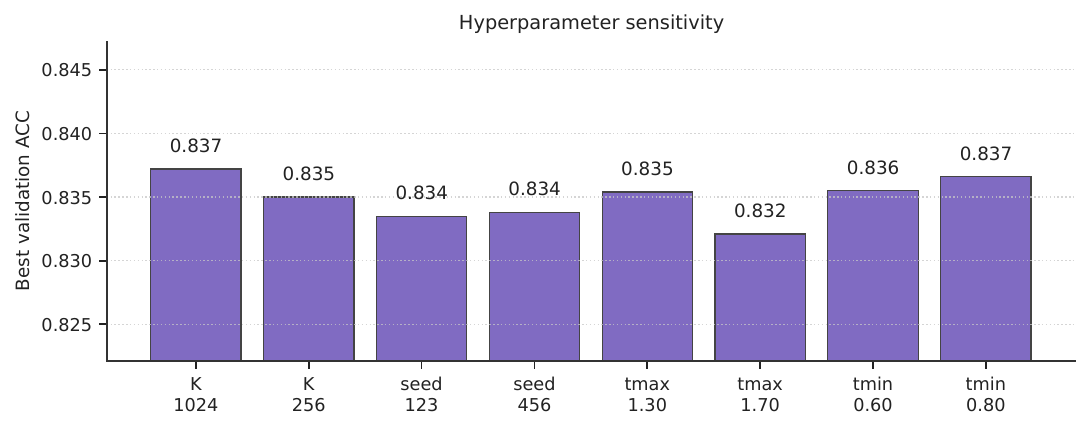}
    \caption{\textcolor{blue}{Hyperparameter sensitivity analysis on Memo2496. The default setting lies in a stable high-performance region across local variations of $\tau_{\min}$, $\tau_{\max}$, queue size $K$, and random seed.}}
    \label{fig:r21_hyper}
\end{figure}

\textcolor{blue}{On Memo2496, Fig.~\ref{fig:r21_convergence} shows smooth convergence and a smaller train-validation gap than the matched baseline, indicating gains without stronger overfitting. The companion analysis in Fig.~\ref{fig:r21_hyper} and Table~\ref{tab:r21_compute_profile} further shows robust local hyperparameter behaviour and reports the additional parameter, time, and memory cost.}

\textcolor{blue}{A temporal sensitivity test on Memo2496 examines whether a fixed excerpt window captures distributed affective evidence. With only part of the $87$-frame sequence visible at inference, full-window evaluation remains strongest, yet the centre half and middle third retain substantial accuracy (arousal: $0.7668$, $0.7498$; valence: $0.6636$, $0.6594$). Together with evidence that musical emotion unfolds over sustained listening \cite{10.1371/journal.pone.0173392,doi:10.1177/0305735611430079,10.1145/2647868.2655019}, this supports a one-minute window whose cues are distributed across the excerpt.}

\begin{figure}[!t]
    \centering
    \includegraphics[width=\linewidth]{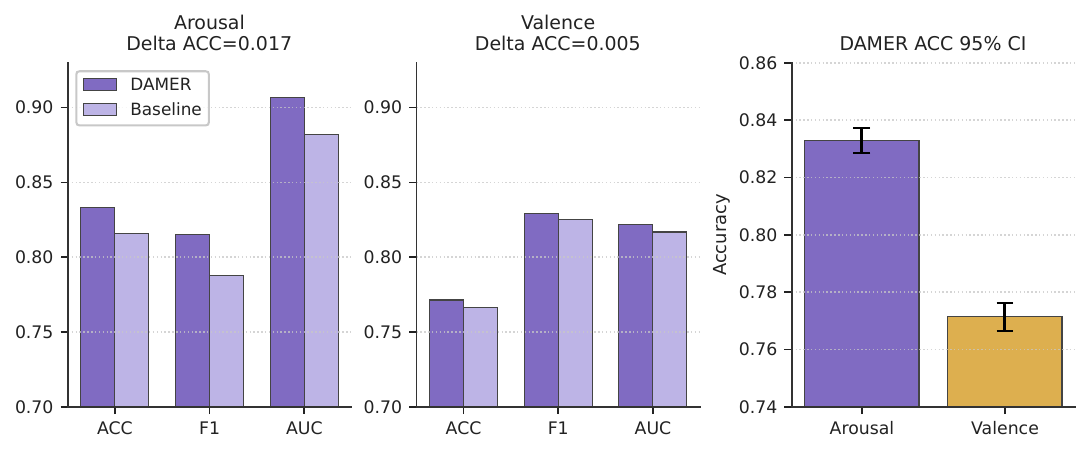}
    \caption{\textcolor{blue}{Paired bootstrap significance analysis on Memo2496 validation data. DAMER remains superior to the matched in-repo ablation baseline in both dimensions, with a smaller margin on valence.}}
    \label{fig:r14_pair}
\end{figure}

\textcolor{blue}{A targeted statistical significance analysis is further conducted on the Memo2496 validation split, where paired sample-level predictions are available under an identical evaluation protocol. To preserve strict pairing, DAMER is compared with an in-repo ablation baseline that removes the cross-view Transformer interaction whilst keeping the remaining training pipeline unchanged. Fig.~\ref{fig:r14_pair} shows that DAMER improves validation accuracy by $0.0172$ on arousal with one-sided paired bootstrap $p < 10^{-4}$, and by $0.0051$ on valence with one-sided paired bootstrap $p = 0.0378$. The corresponding $95\%$ bootstrap confidence intervals for DAMER are $[0.8285, 0.8374]$ for arousal accuracy and $[0.7665, 0.7764]$ for valence accuracy, which supports the claim that the observed gains are stable rather than accidental fluctuations under the matched split.}

\begin{figure}[!t]
    \centering
    \includegraphics[width=\linewidth]{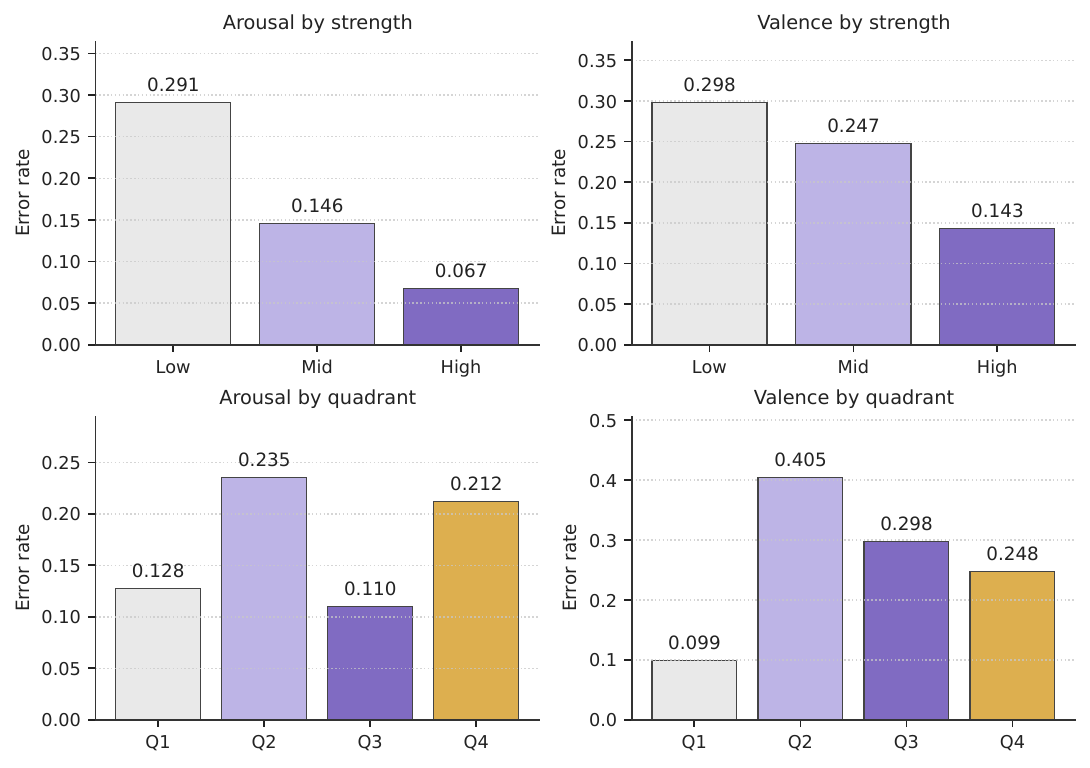}
    \caption{\textcolor{blue}{Difficulty diagnostics on Memo2496 for arousal and valence. Valence shows stronger error concentration in weak emotion regions and in the negative valence and positive arousal quadrant.}}
    \label{fig:r14_difficulty}
\end{figure}

\textcolor{blue}{The same analysis also clarifies why valence remains more difficult than arousal. Fig.~\ref{fig:r14_difficulty} shows that classification error is highest in the weakest emotion strength bin for both dimensions, but the increase is more persistent for valence, where the error rates are $0.2983$, $0.2472$, and $0.1430$ from low to high strength, compared with $0.2912$, $0.1457$, and $0.0674$ for arousal. Quadrant diagnostics further show that the highest valence error occurs in the negative valence and positive arousal region, with an error rate of $0.4046$, whereas the hardest arousal quadrant reaches $0.2355$. These results indicate that DAMER mitigates the asymmetry through cross-view fusion and more reliable supervision, but valence remains intrinsically harder because weak and mixed affective cues are more frequent near the decision boundary.}

\begin{table*}[!t]
\centering
\renewcommand{\arraystretch}{1.08}
\caption{\textcolor{blue}{Compute profile for DAMER and the matched baseline without the cross-view Transformer.}}
\label{tab:r21_compute_profile}
\fontsize{8.5pt}{9.5pt}\selectfont
\begin{tabular}{lccc}
\toprule
Configuration & Parameters (M) & Time per step (ms) & Peak GPU memory (MB) \\
\midrule
DAMER & 1.02 & 66.71 & 1169.79 \\
Baseline without cross-view Transformer & 0.22 & 6.93 & 113.42 \\
\bottomrule
\end{tabular}
\end{table*}

\begin{table*}[!t]
\renewcommand{\arraystretch}{1.04}
\centering
\captionsetup{justification=centering, font=small}
\caption{\textcolor{red}{Ablation classification results on Memo2496, 1000songs, and PMEmo. All values are percentages, and the best metric values within each dataset are shown in bold.}}
\label{tab:ablation}
\fontsize{7.6pt}{8.8pt}\selectfont
\begin{tabular*}{\textwidth}{@{\extracolsep{\fill}}lcccc>{\color{red}}c>{\color{red}}cc>{\color{red}}c>{\color{red}}c@{}}
\toprule
\multirow{2}{*}{\textbf{Dataset}} & \multicolumn{3}{c}{\textbf{Module}} & \multicolumn{3}{c}{\textbf{Arousal}} & \multicolumn{3}{c}{\textbf{Valence}} \\ \cmidrule(lr){2-4} \cmidrule(lr){5-7} \cmidrule(lr){8-10}
 & DSAF & PCL & SAML & ACC & F1 & AUC & ACC & F1 & AUC \\ \midrule
\multirow[t]{7}{*}{\textbf{Memo2496}} & $\checkmark$ & & & 80.94 & 80.09 & 89.22 & 77.40 & 83.45 & 78.91 \\
 & & $\checkmark$ & & 80.58 & 78.45 & 86.85 & 77.34 & 83.42 & 82.11 \\
 & & & $\checkmark$ & 78.38 & 78.25 & 88.27 & 75.79 & 83.79 & 81.96 \\
 & $\checkmark$ & $\checkmark$ & & 81.89 & 79.37 & 87.41 & 77.07 & 82.22 & 77.00 \\
 & & $\checkmark$ & $\checkmark$ & 81.61 & 78.28 & 86.89 & 74.61 & 83.38 & 81.18 \\
 & $\checkmark$ & & $\checkmark$ & 81.08 & 79.07 & 88.20 & 75.59 & 83.29 & 78.54 \\
 & $\checkmark$ & $\checkmark$ & $\checkmark$ & \textbf{82.95} & \textbf{80.70} & \textbf{90.21} & \textbf{78.34} & \textbf{84.72} & \textbf{83.31} \\
\midrule
\multirow[t]{7}{*}{\textbf{1000songs}} & $\checkmark$ & & & 79.02 & 82.06 & 87.42 & 69.51 & 71.30 & 73.90 \\
 & & $\checkmark$ & & 78.69 & 79.42 & 86.89 & 69.42 & 71.72 & 74.76 \\
 & & & $\checkmark$ & 78.14 & 79.84 & 86.81 & 68.46 & 71.85 & 74.53 \\
 & $\checkmark$ & $\checkmark$ & & 80.67 & 81.46 & 89.11 & 70.34 & 72.12 & 74.17 \\
 & & $\checkmark$ & $\checkmark$ & 81.03 & 78.79 & 86.63 & 69.49 & 71.94 & 75.29 \\
 & $\checkmark$ & & $\checkmark$ & 79.36 & 82.23 & 88.40 & 69.55 & 72.21 & 74.02 \\
 & $\checkmark$ & $\checkmark$ & $\checkmark$ & \textbf{81.28} & \textbf{82.32} & \textbf{89.15} & \textbf{70.93} & \textbf{72.79} & \textbf{77.28} \\
\midrule
\multirow[t]{7}{*}{\textbf{PMEmo}} & $\checkmark$ & & & 84.44 & 82.88 & 83.00 & 75.50 & 84.96 & 78.67 \\
 & & $\checkmark$ & & 83.01 & 84.85 & 82.85 & 75.06 & 82.41 & 75.63 \\
 & & & $\checkmark$ & 82.95 & 84.08 & 83.80 & 74.23 & 82.74 & 76.11 \\
 & $\checkmark$ & $\checkmark$ & & 85.78 & 84.18 & 82.92 & 77.10 & 83.28 & 78.08 \\
 & & $\checkmark$ & $\checkmark$ & 83.78 & 84.08 & 83.28 & 76.03 & 82.45 & 75.63 \\
 & $\checkmark$ & & $\checkmark$ & 84.23 & 83.38 & 82.77 & 74.09 & 84.91 & 78.54 \\
 & $\checkmark$ & $\checkmark$ & $\checkmark$ & \textbf{85.98} & \textbf{85.11} & \textbf{84.85} & \textbf{77.61} & \textbf{85.54} & \textbf{79.71} \\
\bottomrule
\end{tabular*}
\end{table*}

\subsection{Ablation Study}
\textcolor{red}{Table~\ref{tab:ablation} reports ACC, positive class F1, and ROC AUC for every ablation configuration. ACC is retained for continuity with the original evaluation, whilst F1 accounts for the precision and recall tradeoff and AUC evaluates ranking performance independently of a single decision threshold. The full DAMER configuration gives the highest value for all three metrics on both target dimensions across Memo2496, 1000songs, and PMEmo. The agreement among ACC, F1, and AUC shows that the observed ablation pattern is not an artefact of accuracy under asymmetric label proportions.}

\begin{figure}[!t]
    \centering
    \includegraphics[width=\linewidth]{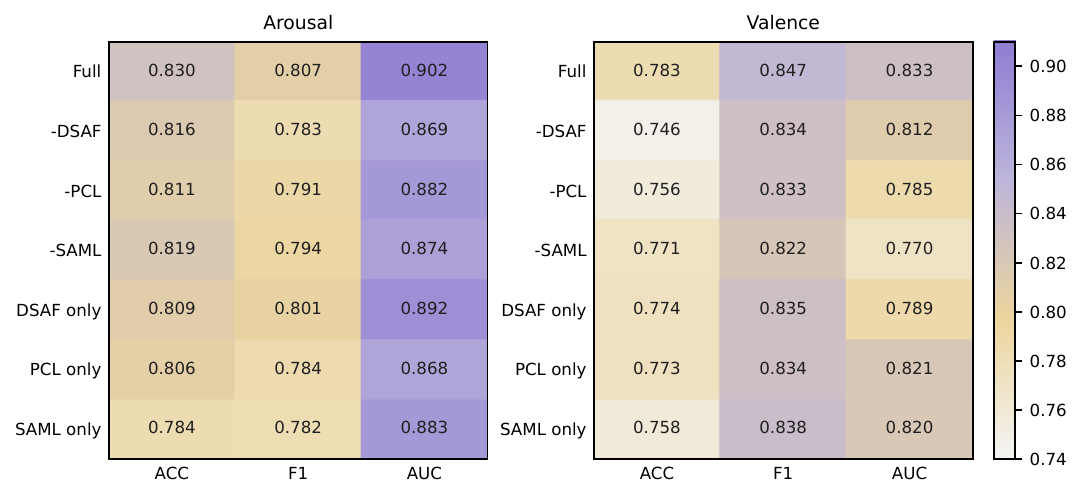}
    \caption{\textcolor{blue}{Ablation heatmap diagnostics on Memo2496.}}
    \label{fig:r12_heatmap}
\end{figure}

\textcolor{red}{The interaction among DSAF, PCL, and SAML combines complementary representation, reliable supervision, and class structure regularisation.} \textcolor{blue}{DSAF addresses the fact that music emotion cues are distributed across complementary acoustic views, so token-level cross-attention improves cue completeness rather than merely increasing feature dimension. PCL is related in spirit to FixMatch \cite{sohn2020fixmatch}, but it is adapted to MER by using cross-view agreement and curriculum temperature control to suppress pseudo-label noise from subjective and locally ambiguous emotion segments.} \textcolor{red}{SAML is related to supervised contrastive learning \cite{khosla2020supervised} and queue-based memory mechanisms such as Momentum Contrast \cite{he2020momentum}. Its labelled queue extends comparison beyond the current mini batch and regularises the dispersion of representations with the same emotion label. Under this formulation, DSAF supplies more complete affective evidence, PCL filters supervision according to that evidence, and SAML encourages a compact emotion class structure.} \textcolor{blue}{Fig.~\ref{fig:r12_heatmap} supports this interpretation by showing that the full DSAF+PCL+SAML configuration maintains the strongest and most balanced profile across ACC, F1, and AUC, whereas single-module settings show narrower strengths and weaker cross-metric stability.}

\begin{figure}[!t]
    \centering
    \includegraphics[width=\linewidth]{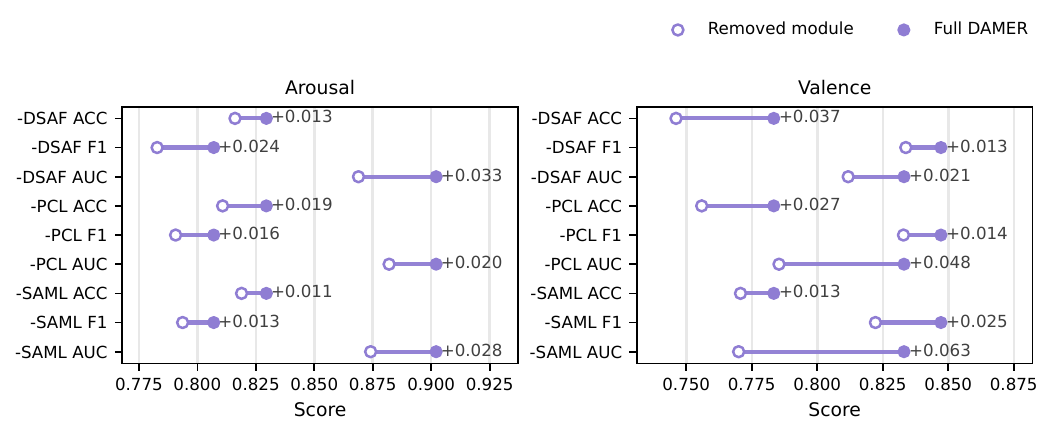}
    \caption{\textcolor{blue}{Dumbbell comparison of full DAMER and remove-one ablation settings on Memo2496.}}
    \label{fig:r12_dumbbell}
\end{figure}

\subsection{\textcolor{blue}{Dual-view Cue Complementarity Analysis}}
\begin{figure}[!t]
    \centering
    \includegraphics[width=\linewidth]{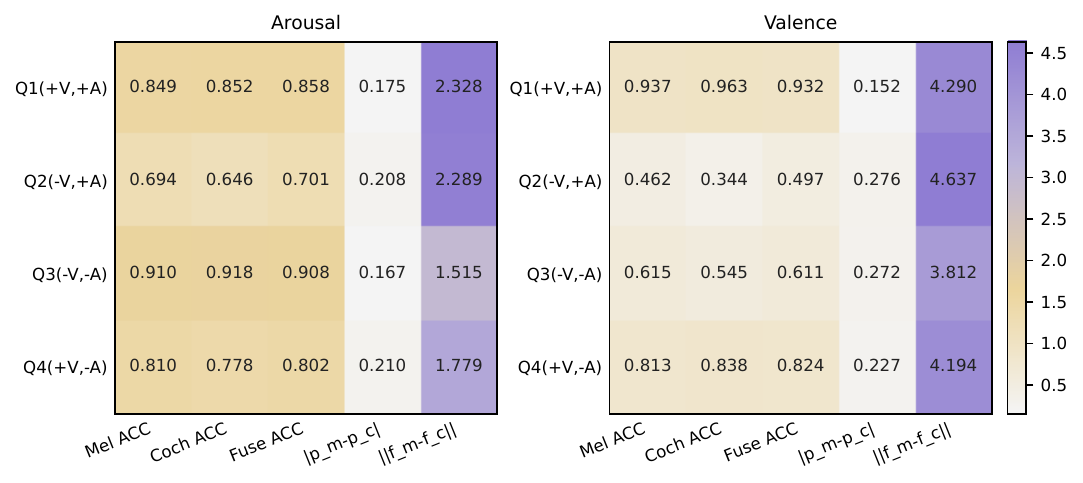}
    \caption{\textcolor{blue}{Quadrant diagnostics for branch and fusion behaviour on Memo2496.}}
    \label{fig:r13_heatmap}
\end{figure}

\begin{figure}[!t]
    \centering
    \includegraphics[width=\linewidth]{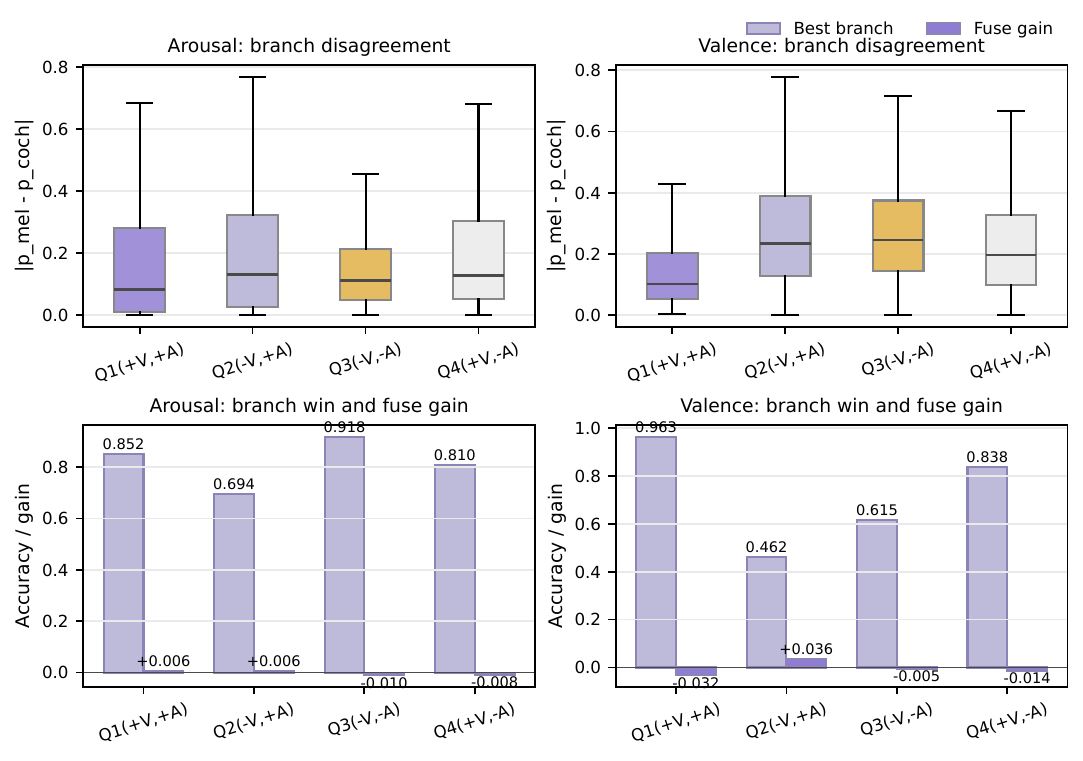}
    \caption{\textcolor{blue}{Branch disagreement and fusion-gain diagnostics by quadrant on Memo2496.}}
    \label{fig:r13_disagree_gain}
\end{figure}

\textcolor{blue}{Fig.~\ref{fig:r13_heatmap} reports quadrant-wise diagnostics for Mel, cochleagram, and fused predictions on Memo2496. In arousal prediction, cochleagram is stronger in Q3, whilst Mel is stronger in Q1 and Q2. In valence prediction, both single branches degrade substantially in Q2 and Q3, and the fused branch provides the most stable recovery in these difficult regions. This pattern indicates that the two views capture different but compatible emotional cues across the valence-arousal space. The companion panels further show that branch disagreement is consistently higher for valence than for arousal, with the highest disagreement in Q2 and Q3, and that positive fuse gain appears most clearly in valence Q2, where fusion improves over the best single branch by a clear margin (Fig.~\ref{fig:r13_disagree_gain}). These observations support a mechanism in which Mel and cochleagram provide complementary evidence, and cross-view fusion is most beneficial when the emotional region is ambiguous for single-view inference.}

\subsection{Progressive Confidence Labelling Dynamics}
\begin{figure}[!t]
    \centering
    \includegraphics[width=\linewidth]{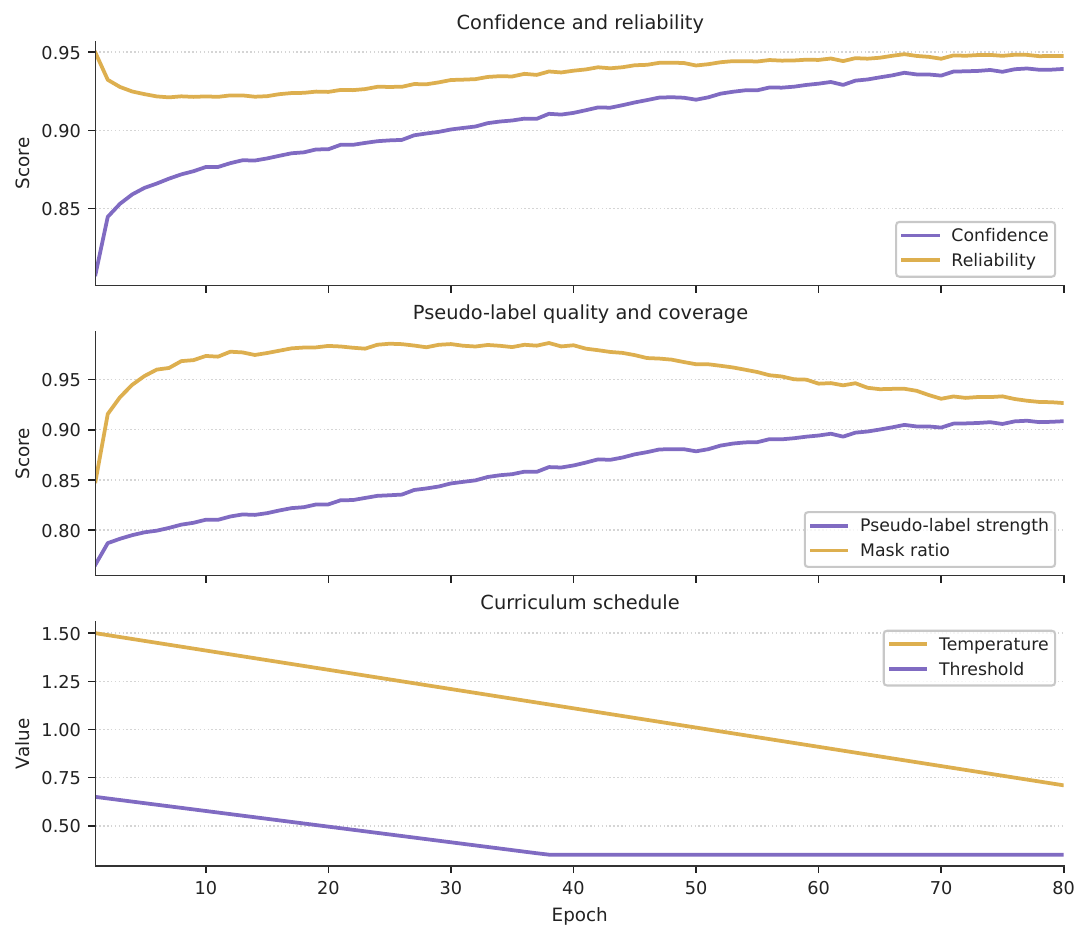}
    \caption{Progressive Confidence Labelling dynamics across the full $80$ training epochs. Top: confidence and reliability score trajectories stabilising near 0.90. Middle: pseudo-label strength and mask ratio demonstrating sustained coverage above $0.92$. Bottom: curriculum schedule showing temperature descent from $1.50$ to $0.71$ and threshold descent from $0.65$ to $0.35$, fully consistent with Eqs.~\eqref{eq9} and~\eqref{eq16}.}
    \label{PCLD}
\end{figure}
Fig.~\ref{PCLD} illustrates the efficacy of the Progressive Confidence Labelling (PCL) module by demonstrating the temporal evolution of curriculum metrics across training epochs. This three-panel visualisation reveals that the DAMER curriculum-based pseudo-labelling strategy maintains consistently high prediction quality whilst progressively refining sample selection criteria. \textcolor{blue}{It reports the complete $80$-epoch training trajectory without truncation in the top or middle panels, as all logged curriculum statistics remain valid throughout training. Diagnostic verification further confirms that the plotted temperature sequence matches Eq.~\eqref{eq9} with zero numerical deviation in the exported statistics.} Confidence and reliability stabilise near 0.90 with diminishing variance, confirming that Jensen-Shannon divergence reliably quantifies dual-branch consistency. Sustained mask ratio and pseudo-label strength validate the strict-to-lenient threshold schedule, whilst concurrent temperature annealing sharpens predictions as $\theta_t$ relaxes to admit ambiguous samples. Architectural coupling reinforces this schedule: DSAF lowers inter-branch divergence to elevate $r=\exp(-\text{JS})$, \textcolor{red}{whilst SAML regularises the dispersion of representations with the same emotion label through supervised contrastive learning.} PCL thus balances coverage and quality without the instability of fixed-threshold pseudo-labelling.

\FloatBarrier
\section{Conclusion}

This paper presents the Dual-view Adaptive Music Emotion Recogniser (DAMER) framework and the Memo2496 dataset for music emotion recognition. Memo2496 is one of the largest expert-annotated instrumental music emotion datasets, featuring $2496$ tracks with continuous Valence and Arousal labels from $30$ specialists. This dataset ensures high reliability and reduced label noise, addressing crowdsourced limitations. DAMER integrates three modules: its Dual-Stream Attention Fusion module leverages complementary acoustic features from Mel spectrograms and cochleagrams via bidirectional cross-attention; its Progressive Confidence Labelling module generates reliable pseudo-labels via curriculum-based temperature scheduling and Jensen-Shannon divergence-based consistency, mitigating confirmation bias; \textcolor{red}{and its Style-Anchored Memory Learning module uses labelled contrastive memory to encourage compact representations among samples with the same emotion label.} \textcolor{red}{Experiments on Memo2496, 1000songs, and PMEmo show that DAMER provides leading performance on Memo2496 and 1000songs and strong valence performance on PMEmo, whilst PMEmo arousal remains competitive across task-specific MER baselines.} Ablation studies, supplementary diagnostic analyses, and cross-dataset evaluation provide evidence for DAMER's efficacy and generalisation. The Memo2496 dataset and DAMER source code are publicly released, facilitating reproducible research.

\section{Acknowledgements}
% The authors express their sincere gratitude to Bianna Chen and Haozhang Yuan, esteemed members of the Data Science and Artificial Intelligence Laboratory (DSAIL) in South China University of Technology, Guangzhou, China. 

\textcolor{blue}{The authors thank the certified music specialists for their careful annotation work and valuable professional input during the construction of the Memo2496 dataset.}
\bibliographystyle{IEEEtran}
\bibliography{IEEEabrv.bib}
\begin{IEEEbiography}[{\includegraphics[width=1in,height=1.25in, clip,keepaspectratio]{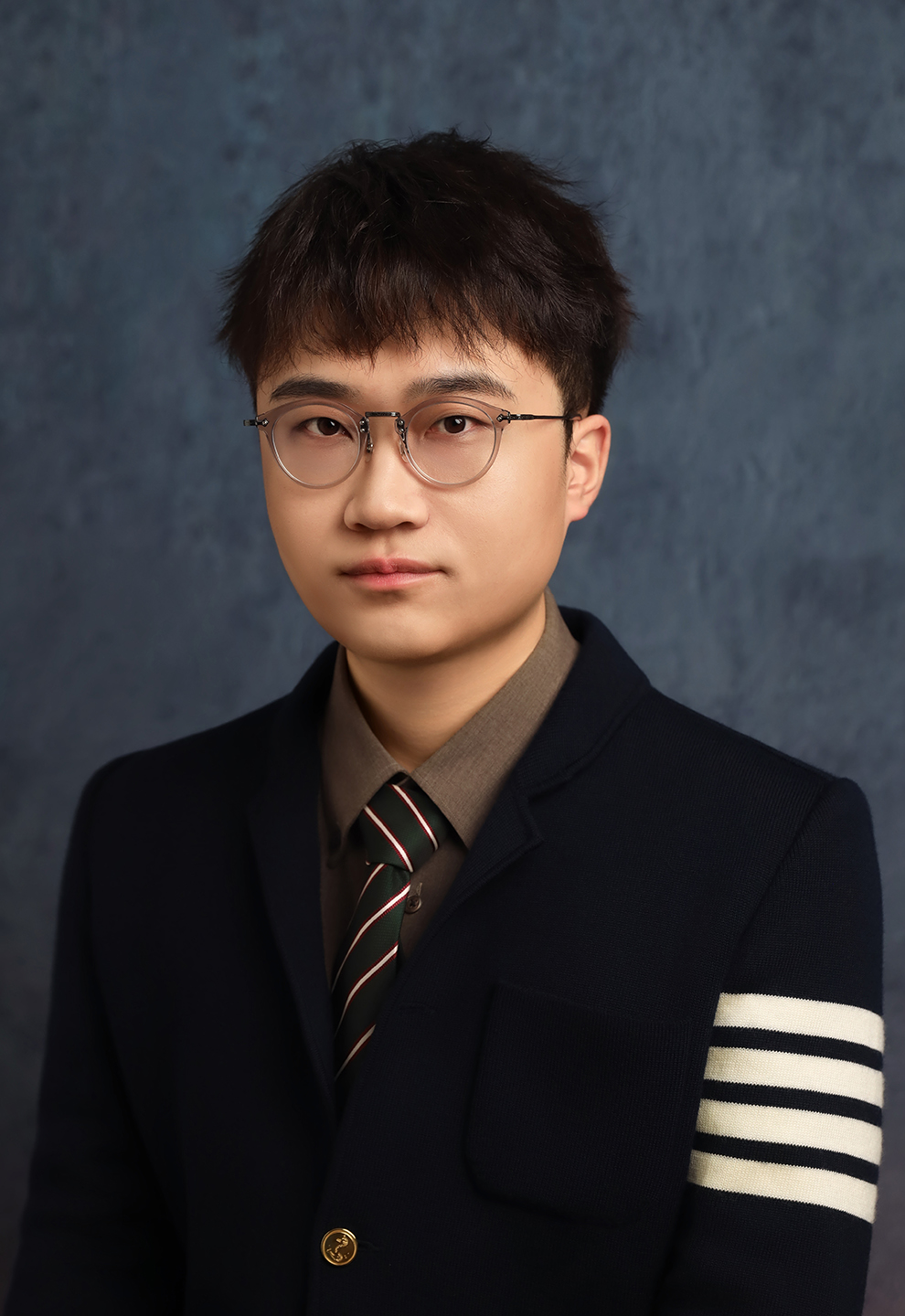}}]{Qilin Li} (Member, IEEE) received the B.Eng. degree in electrical engineering and automation from Wuhan University of Science and Technology, at Wuhan, China, in 2018, the M.S. degree in computer science from University of Wollongong, at Wollongong, NSW, Australia, in 2020 and the Ph.D. degree in computer science and technology from South China University of Technology, at Guangzhou, China, in 2026. He is currently a postdoctoral researcher with the Department of Automation, Tsinghua University, at Beijing, China.
    
His current research interests include affective computing and neural network.	
    \end{IEEEbiography}

%\vspace{-15mm}

\begin{IEEEbiography}[{\includegraphics[width=1in,height=1.25in,clip,keepaspectratio]{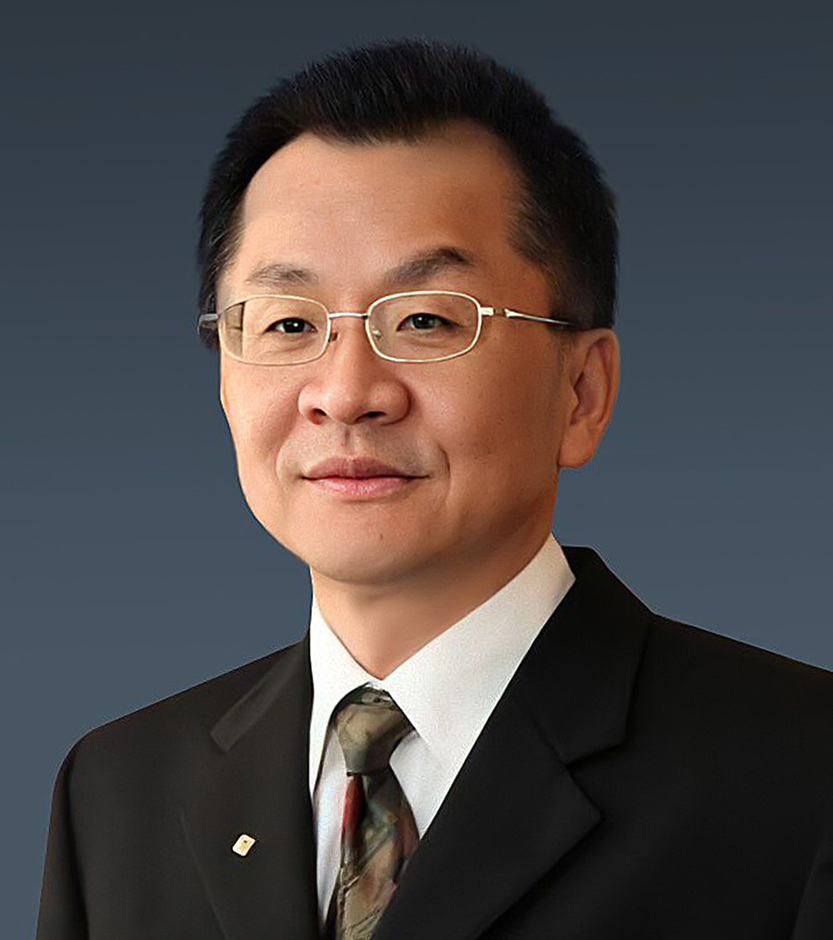}}]{C. L. Philip Chen}  (Life Fellow, IEEE) received the M.S. degree in electrical and computer science from the University of Michigan at Ann Arbor, Ann Arbor, MI, USA, in 1985, and the Ph.D. degree in electrical and computer science from Purdue University, West Lafayette, IN, USA, in 1988.
	
	He is the Chair Professor and the Dean of the School of Computer Science and Engineering, South China University of Technology, Guangzhou, China. Being a Program Evaluator of the Accreditation Board of Engineering and Technology Education in the U.S., for computer engineering, electrical engineering, and software engineering programs, he successfully architects the University of Macau’s Engineering and Computer Science programs receiving accreditations from Washington/Seoul Accord through Hong Kong Institute of Engineers (HKIE), Hong Kong, of which is considered as his utmost contribution in engineering/computer science education for Macau as the former Dean of the Faculty of Science and Technology. His current research interests include cybernetics, systems, and computational intelligence.
	
	Dr. Chen was a recipient of the 2016 Outstanding Electrical and Computer Engineers Award from his alma mater, Purdue University in 1988. He received the IEEE Norbert Wiener Award in 2018 for his contribution in systems and cybernetics, and machine learning. He is also a highly cited researcher by Clarivate Analytics from 2018 to 2023. He is currently the Editor-in-Chief of the IEEE TRANSACTIONS ON CYBERNETICS, an Associate Editor of the IEEE TRANSACTIONS ON ARTIFICIAL INTELLIGENCE, and IEEE TRANSACTIONS ON FUZZY SYSTEMS. He was the IEEE Systems, Man, and Cybernetics Society President from 2012 to 2013, the Editor-in-Chief of the IEEE TRANSACTIONS ON SYSTEMS, MAN, AND CYBERNETICS: SYSTEMS from 2014 to 2019. He was the Chair of TC 9.1 Economic and Business Systems of International Federation of Automatic Control from 2015 to 2017. He is a Fellow of AAAS, IAPR, CAA, and HKIE; a member of Academia Europaea, European Academy of Sciences and Arts.\end{IEEEbiography}

\begin{IEEEbiography}[{\includegraphics[width=1in,height=1.25in,clip,keepaspectratio]{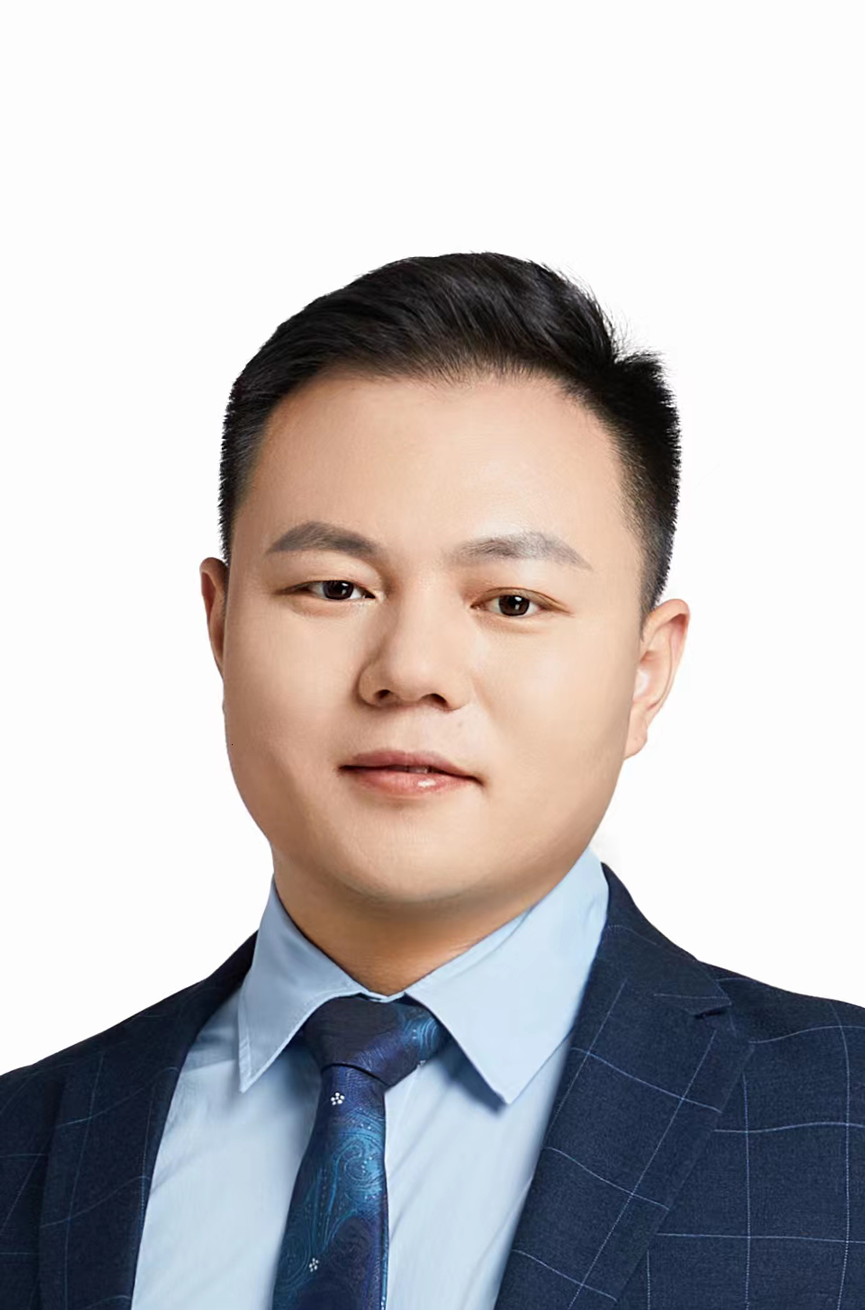}}]{Tong Zhang} (Senior Member, IEEE) received the B.S. degree in software engineering from Sun Yat-sen University, at Guangzhou, China, in 2009, and the M.S. degree in applied mathematics from University of Macau, at Macau, China, in 2011, and the Ph.D. degree in software engineering from the University of Macau, at Macau, China, in 2016. Dr. Zhang currently is a professor and Associate Dean of the School of Computer Science and Engineering, South China University of Technology, China. 

    His research interests include affective computing, evolutionary computation, neural networks, and other machine learning techniques and their applications. Prof. Zhang is the Associate Editor of the IEEE Transactions on Affective Computing, IEEE Transactions on Computational Social Systems, and Journal of Intelligent Manufacturing. He has been working in publication matters for many IEEE conferences. 
        
    \end{IEEEbiography}

%\vspace{-15mm}

%\vspace{-15mm}

\end{document}